\documentclass[final,twocolumn,10 pt,journal]{IEEEtran}

\usepackage{amssymb,amsmath,graphicx,epsfig,cite}
\usepackage[usenames]{color}
\usepackage{soul}
\usepackage[ruled, lined]{algorithm2e}
\usepackage{subfigure}
\usepackage{multirow}
\usepackage{rotating}
\usepackage{tabu}
\usepackage{lscape}
\usepackage{indentfirst}
\usepackage{latexsym}
\usepackage{tabls}
\usepackage{wrapfig}
\usepackage{slashbox}
\usepackage{longtable}
\usepackage{supertabular}
\usepackage{bbm}
\usepackage{amsmath}
\graphicspath{{figures/}}

\newcommand{\comment}[1]{}

\usepackage{ifthen}
\newboolean{isPublishMode}
\setboolean{isPublishMode}{true}

\ifthenelse{\boolean{isPublishMode}}
{
    \newcommand{\PublishMode}[1]{#1} 
    \newcommand{\DraftMode}[1]{}  
}   
{
    \newcommand{\PublishMode}[1]{} 
    \newcommand{\DraftMode}[1]{#1}  
}

\DraftMode{
   \linespread{1.425}  
}

\title{Feasibility Study on Intra-Grid Location Estimation Using Power ENF Signals}

\author{Ravi Garg,~\IEEEmembership{Member,~IEEE}, Adi Hajj-Ahmad,~\IEEEmembership{Member,~IEEE}, and Min Wu,~\IEEEmembership{Fellow,~IEEE}
\thanks{The work shown in this paper was carried out when all the authors were with the Department of Electrical and Computer Engineering, University of Maryland, College Park, MD 20742, USA. Ravi Garg is a Senior Applied Scientist at Amazon (e-mail: ravigarg@amazon.com), Adi Hajj-Ahmad is a Data Scientist at Amazon (e-mail: adihajjahmad@gmail.com), and Min Wu is a Professor at the University of Maryland, College Park (e-mail: minwu@umd.edu). Preliminary version of this work was presented in ICASSP 2013.}
}

\DraftMode{\date{\today}}

\begin{document}

\maketitle
\begin{abstract}
The Electric Network Frequency (ENF) is a signature of power distribution networks that can be captured by multimedia recordings made in areas where there is electrical activity. This has led to an emergence of several forensic applications based on the use of the ENF signature. Examples of such applications include estimating or verifying the time-of-recording of a media signal and inferring the power grid associated with the location in which the media signal was recorded. In this paper, we carry out a feasibility study to examine the possibility of using embedded ENF traces to pinpoint the location-of-recording of a signal within a power grid. In this study, we demonstrate that  it is possible to pinpoint the location-of-recording to a certain geographical resolution using power signal recordings containing strong ENF traces.  To this purpose, a high-passed version of an ENF signal is extracted and it is demonstrated that the correlation between two such signals, extracted from recordings made in different geographical locations within the same grid, decreases as the distance between the recording locations increases. We harness this property of correlation in the ENF signals to propose trilateration based localization methods, which pinpoint the unknown location of a recording while using some known recording locations as anchor locations. We also discuss the challenges that need to be overcome in order to extend this work to using ENF traces in noisier audio/video recordings for such fine localization purposes.
\end{abstract}

\begin{IEEEkeywords}
Electric Network Frequency, Information Forensics, Fine Location-of-recording Estimation
\end{IEEEkeywords}

\section{Introduction}
\label{sec:ch6:intro}
The Electric Network Frequency is the frequency of power distribution networks. It has a nominal value of 60Hz in the United States and Canada, and 50Hz in most other parts of the world. The ENF does not stay at this nominal value, but rather fluctuates around it due to load changes in the power grid. The changing value of the instantaneous ENF over time is known as the \emph{ENF signal}.

An important characteristic of ENF signals, which makes them relevant to multimedia forensics, is that the ENF variations can be captured in media recordings made in places where there is electrical activity. In audio recordings, this is mainly due to electromagnetic influences and the power acoustic hum\cite{Grigoras:2005, Fechner:2014:hummingHum}. In video recordings, this can be attributed to near-invisible flickering of electric lighting \cite{Garg}.

The ENF signals can be extracted from \emph{power recordings} measured using a 1-D signal sampler and recorder, such as a digital audio recorder, connected to a power outlet using a step-down transformer and a voltage divider circuit~\cite{ENF_webpage}. The recorded signal is divided into frames and a frequency estimation algorithm is applied to each frame to determine its dominant frequency in the band near to nominal ENF value~\cite{Ahmad_APSIPA}~\cite{ENF_2020}. The ENF traces present in such power recordings typically has a very high signal-to-noise ratio (SNR), making the extracted ENF signal a high quality reference for the ENF variations. It has been shown that the ENF variations extracted from audio/video recordings match with the ENF variations from power recordings made simultaneously and within the same power grid~\cite{Garg}.

The similarity in the observations between ENF signals extracted from simultaneously recorded signals has motivated one of the early proposed ENF-based forensics applications: using the ENF traces to authenticate or identify the time-of-recording of a signal~\cite{Grigoras:2005}. Other proposed ENF-based applications include detection of tampering/modification in a media signal~\cite{ENF_spectral_jour, Rodrigues:2013:authenticity, Esquef:2014:authenticity, Hua:2016}~\cite{Nasir_ENF}, multimedia synchronization~\cite{Su:2014:synch, Su:2014:acm:synch}, characterizing the video camera producing an ENF-containing video~\cite{HajjAhmad:SPL:2016}, and determining the location-of-recording among different grids~\cite{HajjAhmad:2013:WIFS,HajjAhmad:2015:tifs:location}. Being able to tell the grid-location of recording of a media signal containing ENF traces across grids naturally leads to the question of whether we can pinpoint the location of recording of the media signal within a grid, i.e., an intra-grid localization. An answer in the affirmative could potentially lead to many applications, especially in defense and security.

At the intra-grid level, most existing works have assumed that ENF signals across different locations in an interconnected power grid are the same at a given time. However, minor variations are likely to be present in the frequency fluctuations at different locations. This can be attributed to local changes in the load and the finite propagation speed of the effects that such load changes would have on other parts of the grid~\cite{ENF_propagation}. In the past, research has been conducted on building a robust reference ENF database for forensics matching by estimating big system disturbances in concurrent multiple recordings at different geographical locations, and using the recordings without any frequency glitch to record ENF at corresponding time~\cite{ENF_disturbances_robust}. Recently concurrent to the review of this journal paper, a machine learning based classifier approach was used to statistically learn the signal patterns unique to a given location, and determine the location out of the possible locations present in the training set~\cite{Tenesse_ML}. However, this classification based method is not designed to infer any location outside those in the training set, and would require gathering new data and retraining the classifier for each new location of interest.

In this paper, we study the variations present in the ENF signals in a grid by conducting experiments on power ENF data collected from several locations within the Eastern US and the Western US grid. We show that differences exist among ENF signals extracted from recordings made simultaneously in various locations within the same interconnected power grid. These differences are more pronounced in the high frequency details of the ENF signals, and can be extracted using a signal processing filtering mechanism. We demonstrate a close relationship between the correlations of the ENF signal details at different locations and the corresponding geographical distances between them in a densely populated part of the grid.

Based on the correlation-distance relationship, we propose two trilateration based localization methods to estimate the locations-of-recordings within the same grid. Our proposed localization methods are scalable in the sense that they can be used to estimate location of recording anywhere in the grid, with concurrent data recordings obtained from a few anchor locations. The correlation-distance relationship can be further rectified and localization accuracy improved by adding data from more locations and at different scales, especially in sparsely populated regions in the power grid. This is in contrast to the machine learning classification system proposed in~\cite{Tenesse_ML}, which can determine the locations out of the $N$ possible ones from the training, but does not provide information about any other locations unforeseen by the training. To the best of our knowledge, this paper together with its preliminary conference version~\cite{Ravi_Localization}, is the first line of work addressing fine intra-grid ENF-based localization of recordings without any need of a reference or training recording obtained from the query location.

In order to understand the feasibility and challenges of localization via the ENF signals extracted from multimedia recordings, we further perform a sensitivity analysis on the effect of noise. Our analysis reveals that a high signal-to-noise ratio~(SNR) is needed to localize media recordings using the current ENF extraction techniques. Nevertheless, such capabilities of the ENF signals may become possible in the future, with a breakthrough improvements in the ENF extraction techniques that will improve the SNR of the ENF signals from media recordings. We also discuss an emerging research direction tailored to security of connected Internet of Things~(IoT) and Cyber Physical Systems~(CPS), using ENF location signatures to facilitate the integrity authentication of sensing~\cite{Wu:2017:Sensys}.

The rest of this paper is organized as follows. Section~\ref{sec:ch6:ENFPropagation} explains the propagation mechanism of the ENF signal, providing intuitions for the work presented in this paper. Section~\ref{sec:ch6:locDependence} discusses different case studies to examine the location dependence of ENF signals extracted from recordings made in the same grid. Section~\ref{sec:ch6:methods} explores two methods for ENF-based fine localization of power recordings within a grid. Section~\ref{sec:ch6:sensitivity} provides further discussions about applications of the fine localization capabilities of ENF signals in media recordings and connected IoT and CPS systems, and Section~\ref{sec:ch6:discuss} concludes the paper.

\section{Background: Propagation Mechanism of ENF}
\label{sec:ch6:ENFPropagation}
The fluctuations in ENF signals in the same grid are due to the dynamic nature of the grid load. Power demand and supply in a given area often follow approximately a cyclic pattern. For example, the demand increases during evening hours in a residential neighborhood, as people switch on lights, cooling/heating systems and other power equipments. For robust operation of the grid, any load change within a grid is regulated~\cite{power_book}. An increase in the load causes the supply frequency to drop temporarily; the control mechanism senses the frequency drop and starts supplying increasing power from adjoining areas to compensate for the increased demand. As a result, the load in adjoining areas also increases, which leads to a drop in the instantaneous supply frequency. The overall power supply will elevate to compensate for the rising load, which leads to a drop in the instantaneous supply frequency in those regions. A similar mechanism compensates for an excess supply of power flow that leads to surges in the supply frequency.

A small change in the load in a given area may have a localized effect on the ENF in that area. However, a large change such as the one caused by a major generator failure may have a substantial effect on an entire grid. In the Eastern US grid, these changes are shown to propagate along the grid at a speed of approximately 500~miles/second~\cite{ENF_propagation} although no major intra-location differences were suggested in absence of such catastrophic events. We conjecture that load change may introduce location specific signatures in the ENF patterns, and such differences, despite being small, may be exploited to narrow down the location of a recording within the grid. Each location sees an aggregation of the effect both generated locally and propagated from many other places in the grid. As a result of such aggregations, we anticipate that ENF signals could have greater similarity for locations close to one another than those further apart. If this conjecture can be verified, to the best of our knowledge, it may suggest, for the first time to the technology community, that there is a potential to pinpoint the location-of-recording within a grid by comparing the similarity of ``microscopic traces" in the ENF signal in question with ENF databases that may be available for a set of anchor locations within that grid.

\begin{figure}
\begin{center}
\subfigure[Normal view]{\includegraphics[width=.32\textwidth]{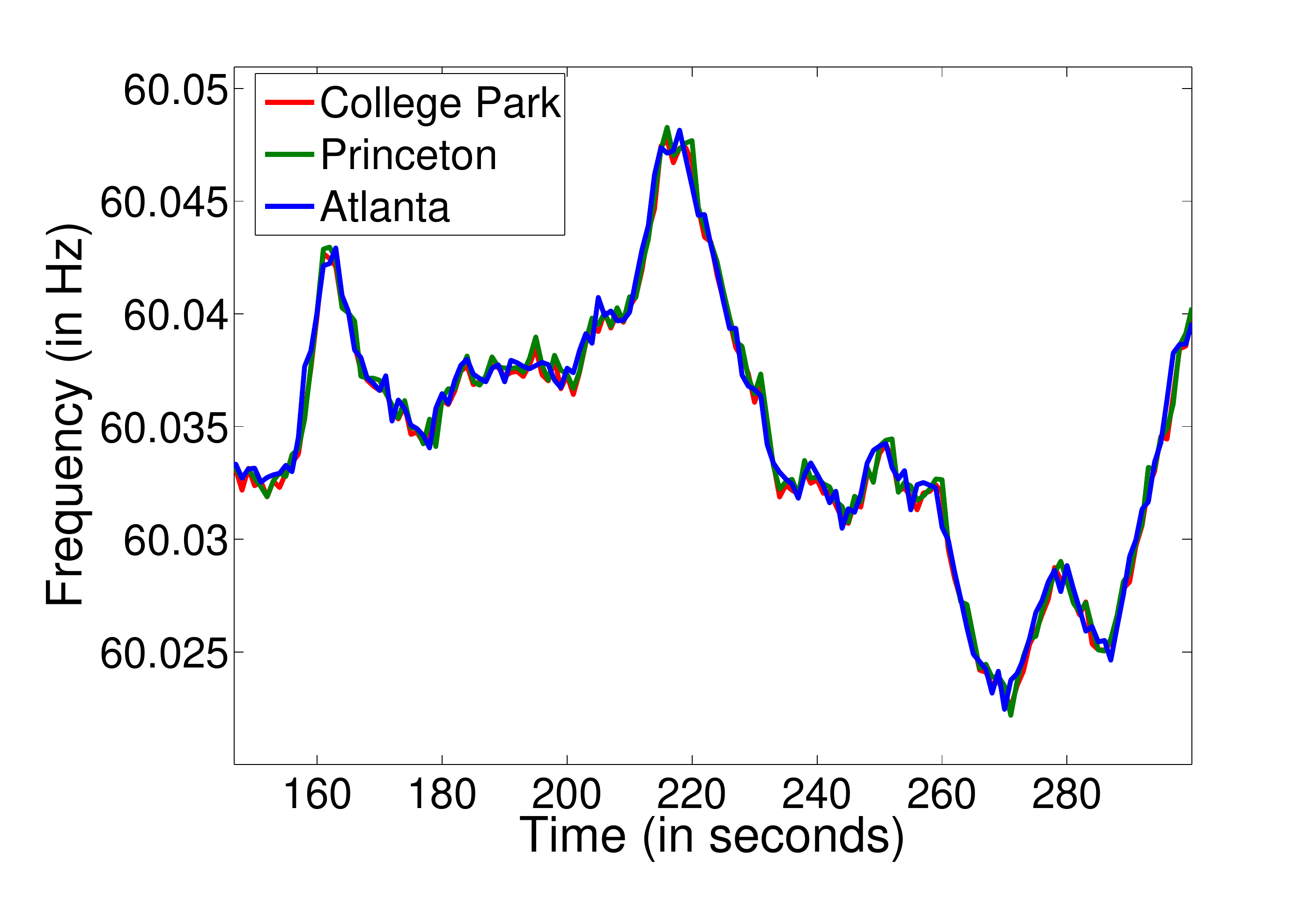}}
\subfigure[Zoom view around 200-230 seconds]{\includegraphics[width=.32\textwidth]{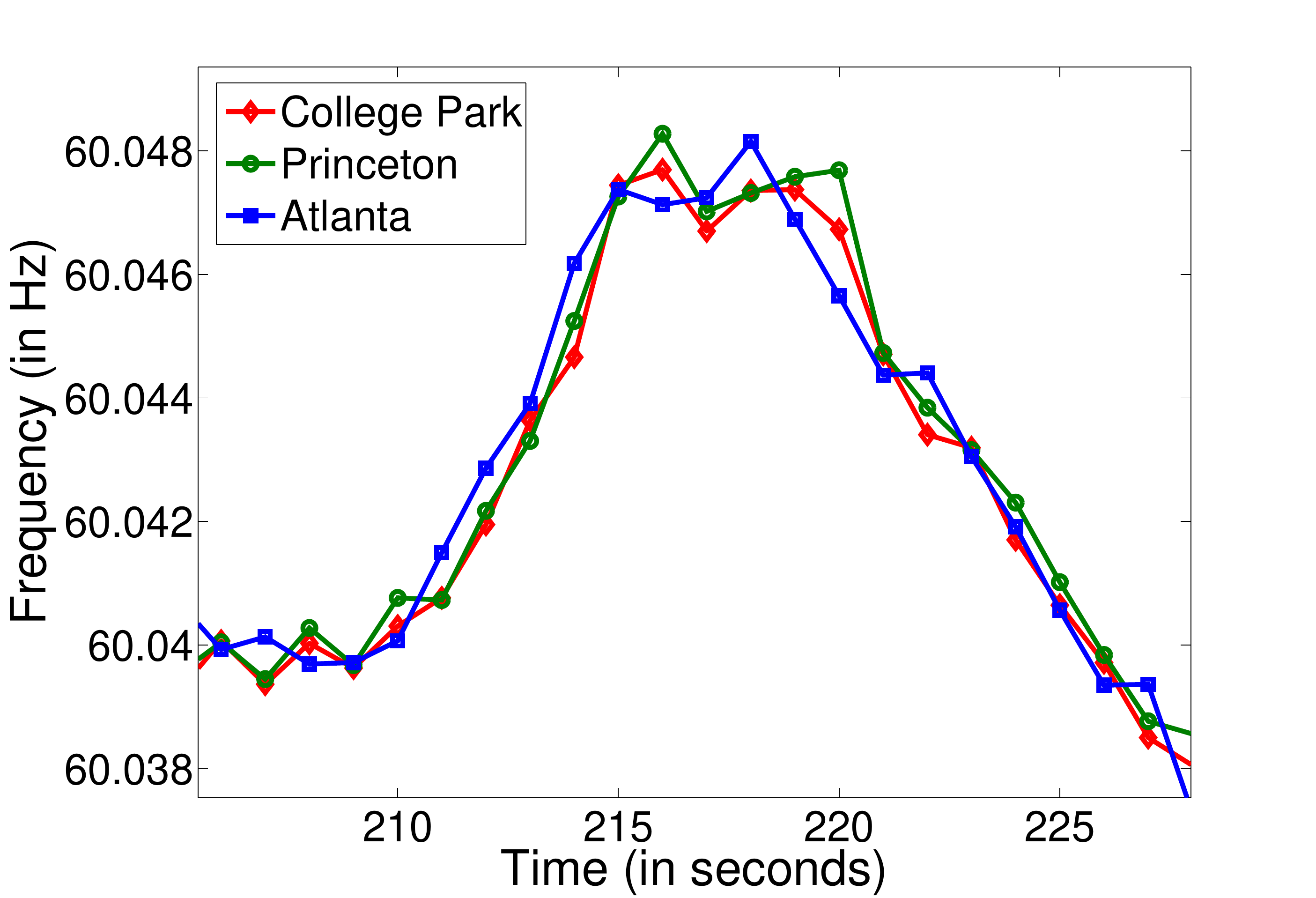}}
\end{center}
\caption[Sample ENF signals from three location recordings in the Eastern US grid]{Sample ENF signals extracted from recordings made in three locations in the Eastern US grid at the same time. (Figures are best viewed in color).}
\label{fig:ENF3Locations}
\end{figure}

\section{Location Dependence of ENF Signals}
\label{sec:ch6:locDependence}
As a first step to explore the availability of location dependent properties of ENF signals, we focus on the ENF signal obtained directly from the power mains. This provides the most favorable conditions in terms of a high signal-to-noise ratio (SNR) of the power ENF signal. ENF signals collected across different locations within the same power grid are similar to each other over time. Following this, an exploration using high SNR signals can enhance the understanding of whether or not ENF signals exhibit location specific characteristics that can be exploited towards building a localization protocol. Such a study is a building block towards solutions to the more difficult problem of location estimation from audio and video recordings, where ENF traces in such recordings may be present at a low SNR and possibly in a distorted form.

Figure~\ref{fig:ENF3Locations}(a) shows a plot of ENF signals extracted from three simultaneous short recordings carried out in College Park in Maryland, Princeton in New Jersey, and Atlanta in Georgia. These three areas are located in the Eastern US grid. The ENF signals are extracted using the subspace analysis based Multiple Signal Classification (MUSIC) algorithm, with instantaneous ENF measured using non-overlapped frame length of 1-second each~\cite{Ahmad_APSIPA}. From Figure~\ref{fig:ENF3Locations}(a), we observe that all three ENF signals correlate strongly at a macroscopic level. In the enlarged plot shown in Figure~\ref{fig:ENF3Locations}(b), however, differences become noticeable across the three recordings. We extract these variations using a filtering mechanism, and then compare them to gain an understanding of the relationship between signals recorded at different locations.

\subsection{Signal Processing Mechanism to Extract ENF Variations}
\label{subsec:ch6:ENFSignalProc}
As seen in Figure~\ref{fig:ENF3Locations}(b), variations occur at high frequencies in ENF signals extracted from simultaneously recorded signals made across different locations of the same grid. To extract these variations, we temporally align the extracted ENF signals by estimating the correlation coefficients between any two signals at different time-lags and finding the time-lag corresponding to the peak value of correlation coefficients~\cite{Garg}. We then use a high pass filtering mechanism in which we pass the temporally aligned ENF signal $f^{\{k\}}(n)$ recorded at $k^{th}$ location through a smoothening filter, and subtract the resulting output signal from $f^{\{k\}}(n)$. The corresponding high pass filtered output, $f_{hp}^{\{k\}}(n)$ is given by:
\begin{equation}
\label{eq:filtering}
f_{hp}^{\{k\}}(n)  = f^{\{k\}}(n)-\sum_{m=-\frac{M-1}{2}}^{\frac{M-1}{2}}w(m)f^{\{k\}}(n-m),
\end{equation}
where $f^{\{k\}}(n)$ is the ENF value at time $n$, $w(\cdot)$ is the coefficient of the smoothening filter, and $M$ is the filter order for feature extraction, chosen as an odd number. We use a very simple smoothetning filter with each coefficient being the inverse of the filter length $M$, and can be represented as $w(m)=\frac{1}{M}$ for $m=-\frac{M-1}{2},..., \frac{M-1}{2}.$

\begin{figure}
\begin{center}
\subfigure{\includegraphics[width=.49\textwidth]{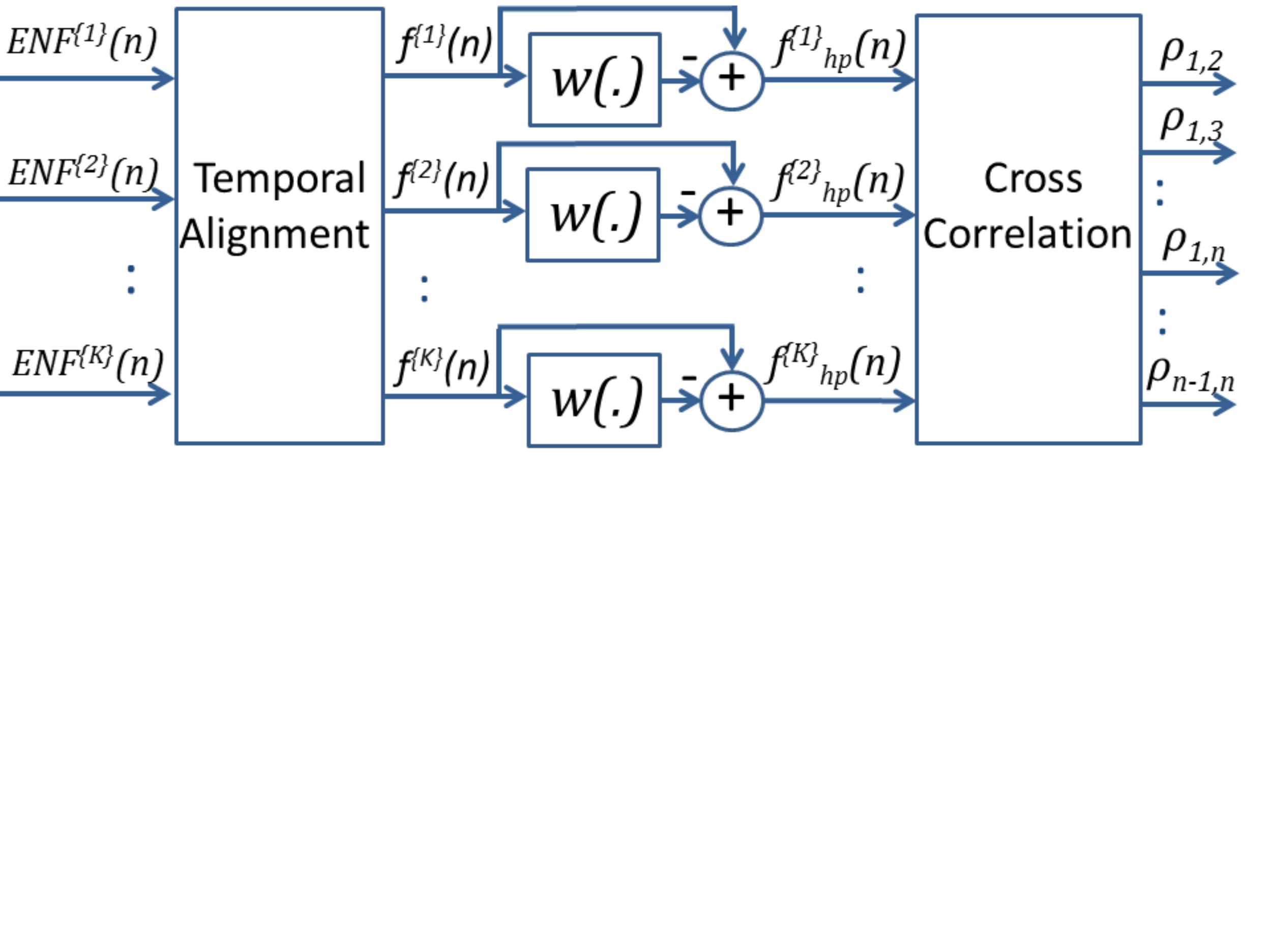}}\vspace{-38mm}
\end{center}
\caption[Signal processing mechanism to extract intra-grid ENF Signatures]{Signal processing mechanism to extract intra-grid ENF Signatures.}
\label{fig:mechansim}
\end{figure}

After extracting high pass filtered signals for each location, their pair-wise cross-correlation coefficients are obtained. The pair-wise cross-correlation coefficient between any two filtered segments at time $n$ from the $k^{th}$ and the $l^{th}$ location is given by:
\begin{equation}
\label{eq:filtering}
\rho_{k,l} = \frac{\sum_{p=0}^{N-1}f_{hp}^{\{k\}}(n+p)f_{hp}^{\{l\}}(n+p)}{\sqrt{\sum_{p=0}^{N-1}(f_{hp}^{\{k\}}(n+p))^2}\sqrt{\sum_{p=0}^{N-1}(f_{hp}^{\{l\}}(n+p))^2}},
\end{equation}
\noindent where $N$ is the length of the signal segment. A block diagram representing this signal processing mechanism is shown in Figure~\ref{fig:mechansim}.

\subsection{Case Study 1: Location Traces from the Eastern and Western US Grid}
\label{subsec:ch6:3Locations}
\begin{figure}
\begin{center}
\subfigure{\includegraphics[width=.48\textwidth]{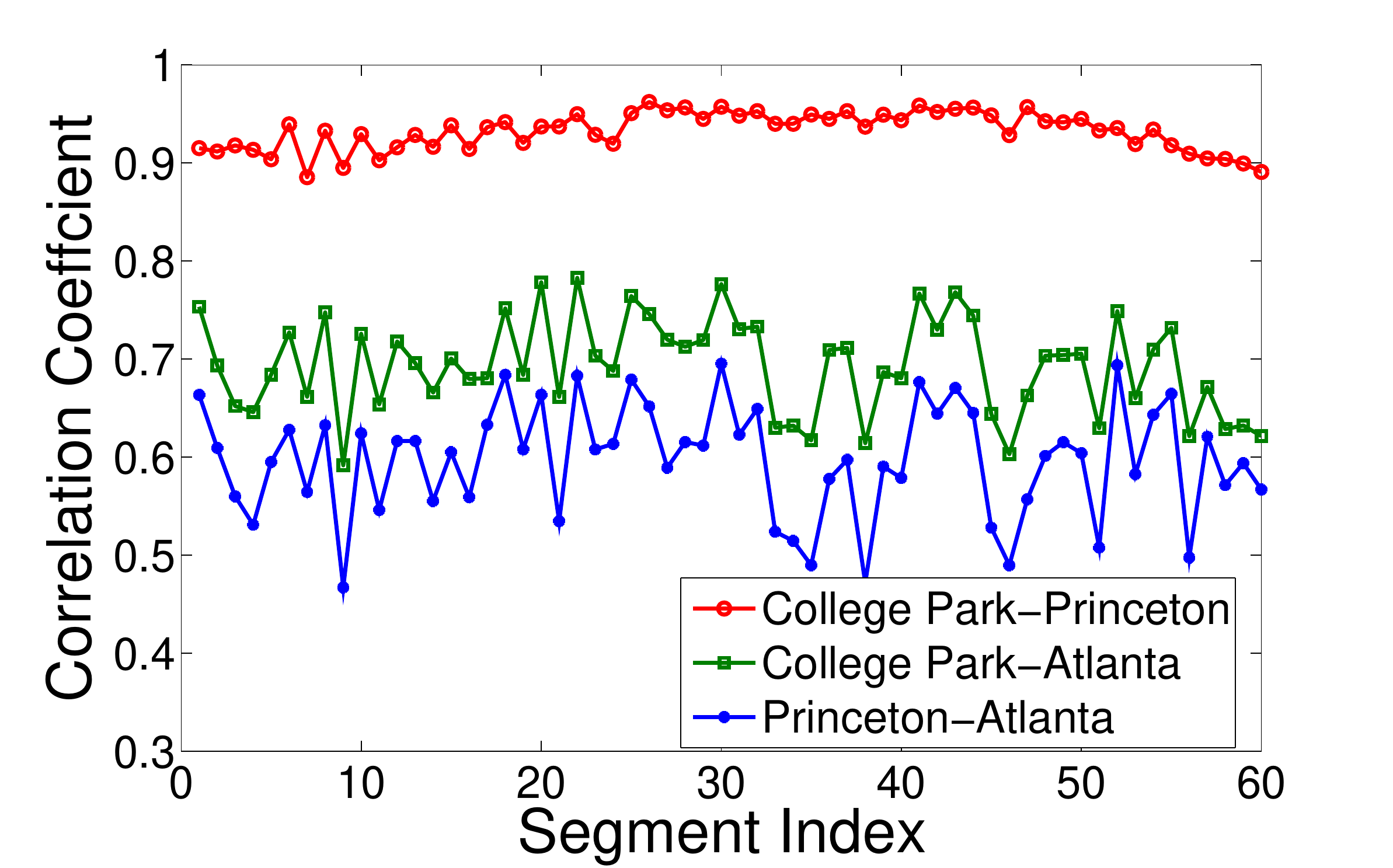}}
\end{center}
\caption[Correlation coefficient between processed ENF signals for 3-location data in US East Coast]{Correlation coefficient between processed ENF signals for 3-location data in the US East Coast for a 10-minute long query segment.}
\label{fig:ENF3LocationCorrelation}
\end{figure}

\begin{figure}
\begin{center}
\includegraphics[width = 0.4\textwidth]{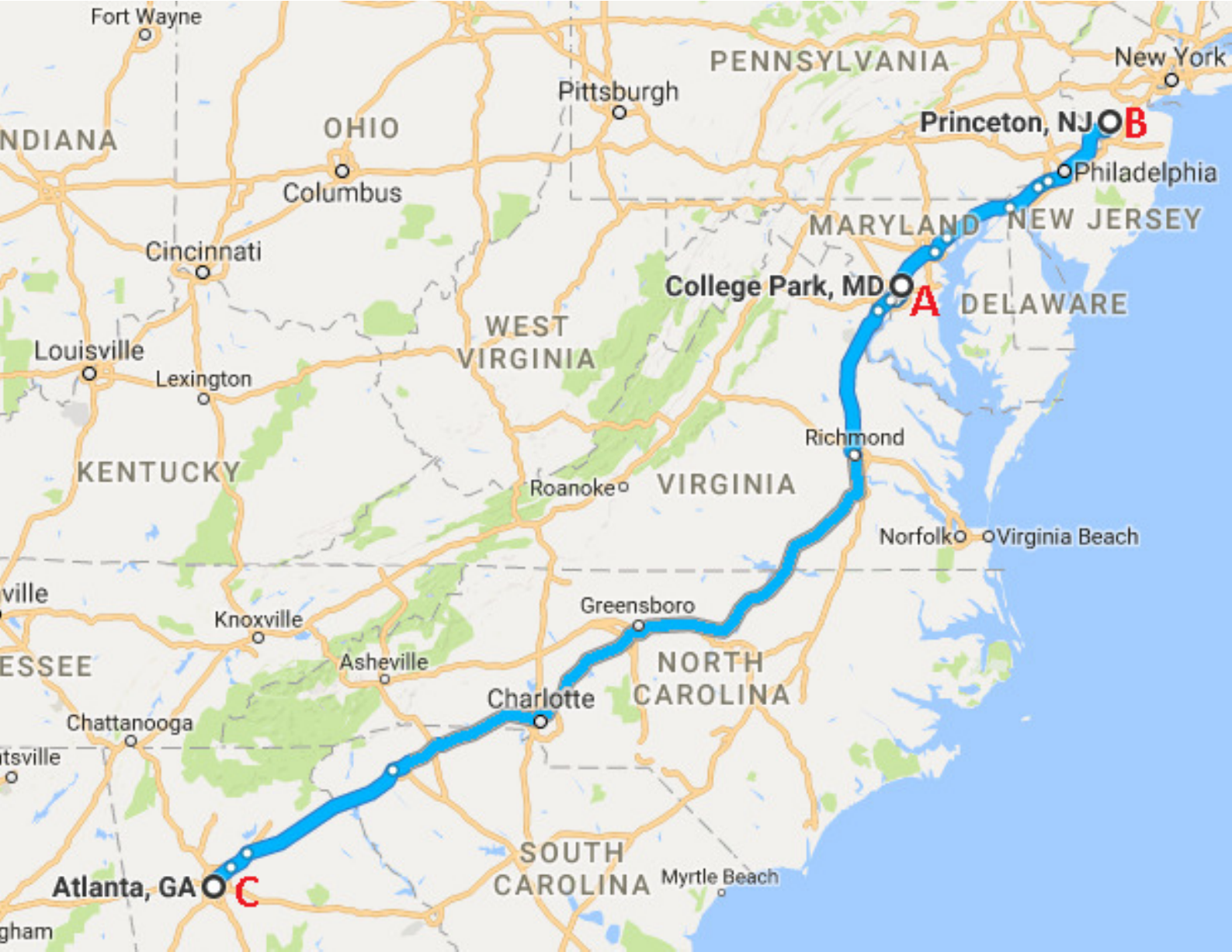}
\end{center}
\caption[Three locations shown on a map for Case Study 1]{Three locations shown on a map for Case Study 1: (A) College Park, MD (City 2), (B) Princeton, NJ (City 1), and (C)Atlanta, GA (City 3). (Best viewed in color)}
\label{fig:3LocMap}
\end{figure}

\paragraph{Correlation in US east grid}
In this section, we describe our experiments on a set of 10-hour long simultaneous recordings of power data from three locations in the Eastern US grid: College Park in Maryland, Princeton in New Jersey, and Atlanta in Georgia. We use the mechanism described in Section~\ref{subsec:ch6:ENFSignalProc} to estimate the cross-correlation between filtered ENF data from all three locations. We divide the signal into non-overlapping segments of length 10~minutes each. Instantaneous frequency (ENF) estimates are obtained every 1~second using the subspace based MUSIC~\cite{freq_book} method, which can provide better frequency estimation accuracy than other methods~\cite{Ahmad_APSIPA}. The plot of the correlation coefficients between processed ENF signals at different locations for filter order $M=$~3 is shown in Figure~\ref{fig:ENF3LocationCorrelation}. This figure shows that the correlation coefficient between the signals from city pairs farther apart in geographical distance is less than that between the signals from the closer city pairs. The correlation coefficient is approximately proportional to the distance between the cities. These three cities lie approximately on a straight line on a map, as shown in Figure~\ref{fig:3LocMap}.

\begin{figure}
\begin{center}
\subfigure{\includegraphics[width=.48\textwidth]{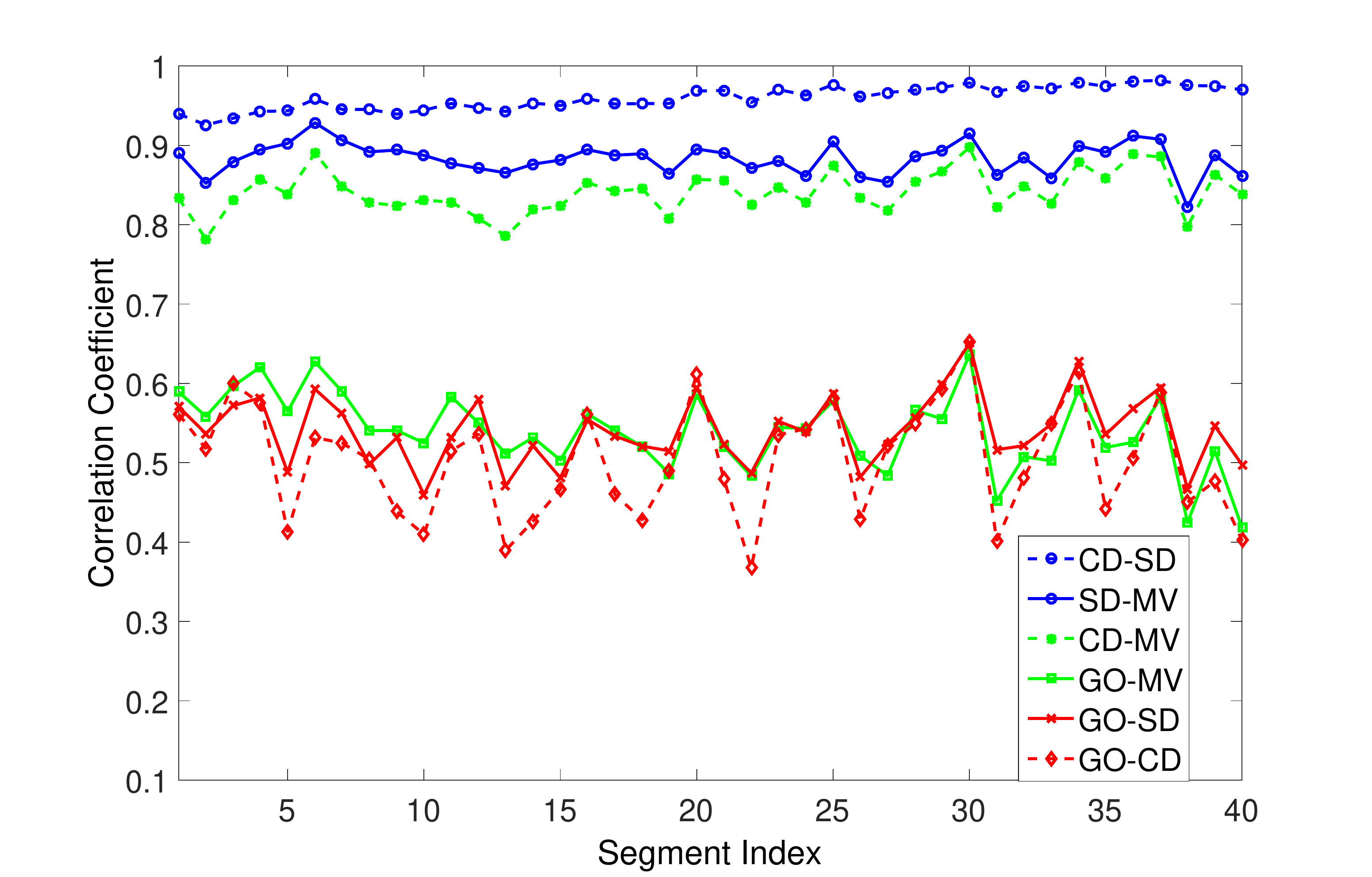}}
\end{center}
\caption[Correlation coefficient between processed ENF signals for 4-location data in US west grid]{Correlation coefficient between processed ENF signals for 4-location data in the US west grid for a 10-minute long query segment.}
\label{fig:CorrCoef_West}
\end{figure}

\begin{figure}
\begin{center}
\subfigure{\includegraphics[width=.48\textwidth]{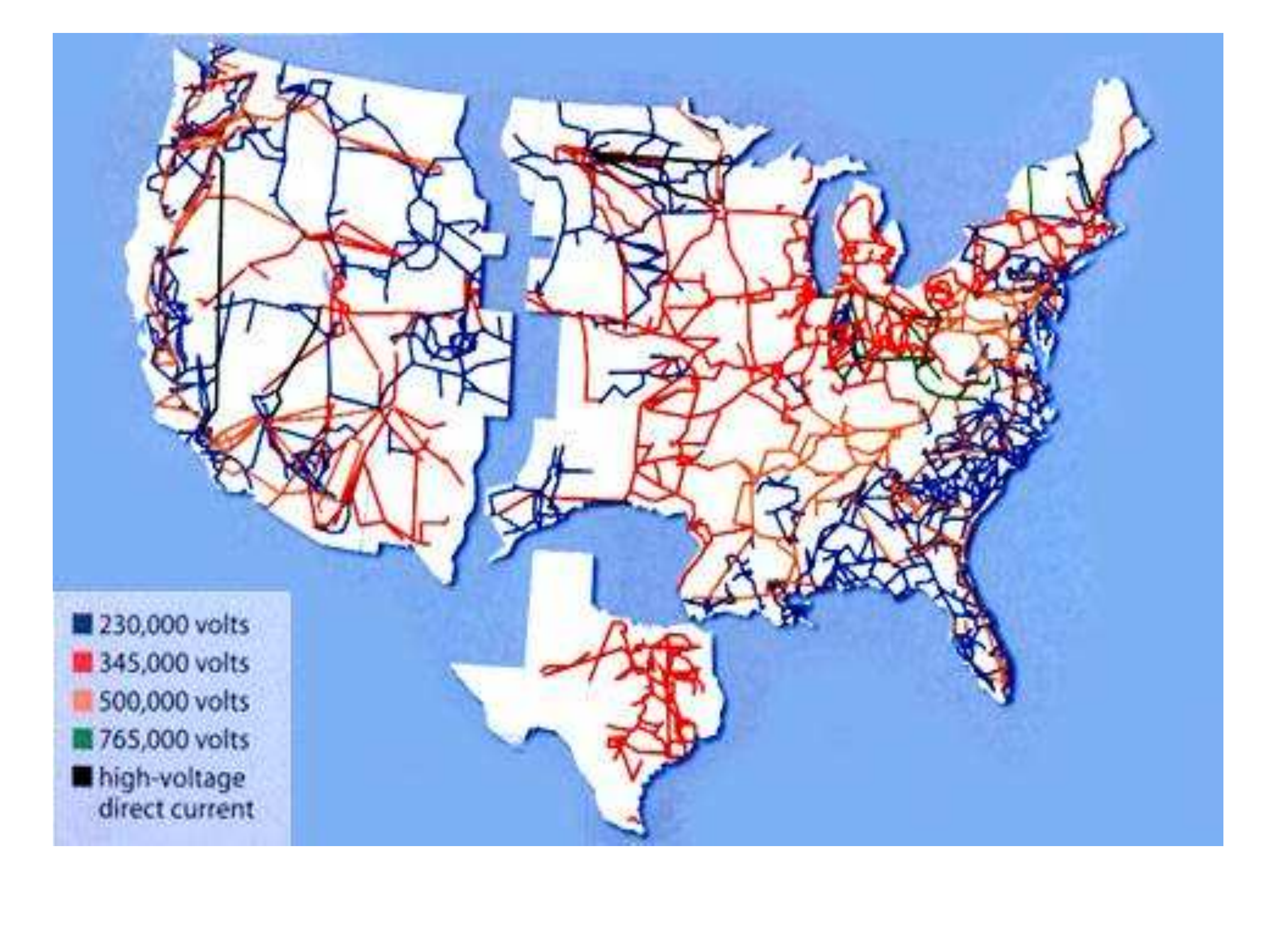}}
\end{center}
\caption[Grid density of the Eastern US interconnection grid]{Grid density of the Eastern US interconnection grid. [\footnotesize{Source: http://www.slipperybrick.com/2009/03/worm-virus-could-bring-down-us-power-grid/}]}
\label{fig:GridDensityUS}
\end{figure}

\paragraph{Correlation in US west grid} To further verify the consistency of the correlation coefficient - distance relationship in another power grid, we also conduct a concurrent power mains recording at 4-locations in the US west grid. The four locations are chosen at densely populated regions of Mountain View (MV) in California, San Diego (SD) in California, Chandler (CD) in Arizona, and sparsely populated Golden (GO) in Colorado. The plot of correlation coefficient for the same settings used in 3-location US east coast recordings is shown in Figure~\ref{fig:CorrCoef_West}. From this figure, we observe a similar correlation v.s. distance relationship to the US east coast, especially for densely populated region of the grid. The value of correlation coefficient is highest for CD-SD correlation, which is closest in terms of geographical distance among the four locations. The correlation coefficient decreases for SD-MV recording as the distance between these cities is more as compared to CD-SD distance. Golden in Colorado is the remotest of all locations, and a close examination of the US grids density map shown in Figure~\ref{fig:GridDensityUS} reveals that while grid connection between Denver region (main metropolis near Golden) and Mountain View, CA is on a straight line route, the connection to SD and CD is not a straight line point-to-point connect, possibly because of mountainous terrain. Due to its far location from remaining 3 locations in our dataset, Golden exhibits much smaller value of correlation coefficient with the 3 locations, as compared with other location pairs. Based on these observations, we derive a relationship between the correlation coefficients among the data from different locations and the corresponding geographical distances. To avoid redundant exploration, we will present our follow-up analysis in the remaining part of the paper on the US east coast recordings.

\begin{figure}
\begin{center}
\subfigure[For different M, ENF estimated at 1-sec durations ]{\includegraphics[width=.32\textwidth]{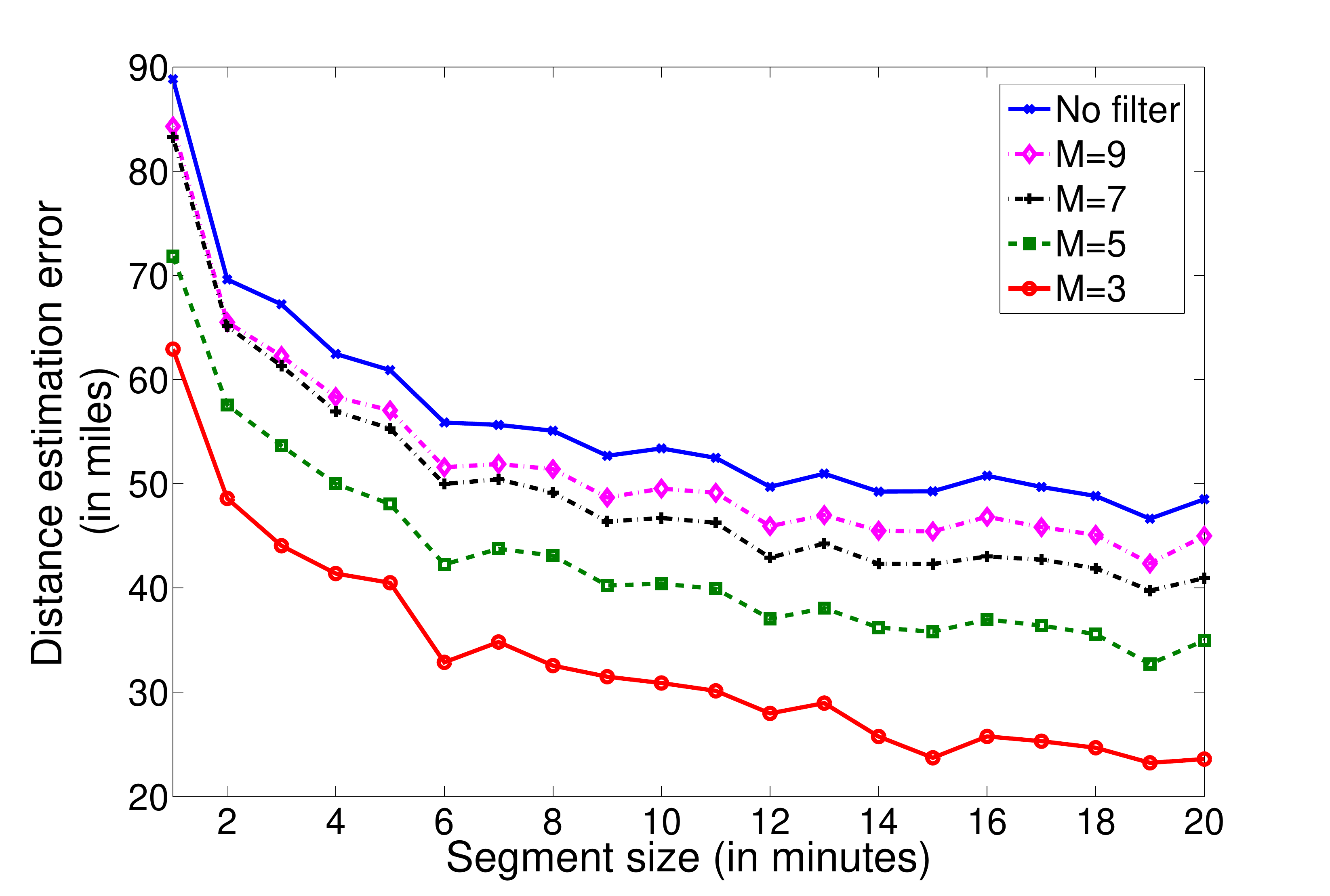}}
\subfigure[For different durations for instantaneous frequency (ENF) estimation, $M=3$ ]{\includegraphics[width=.32\textwidth]{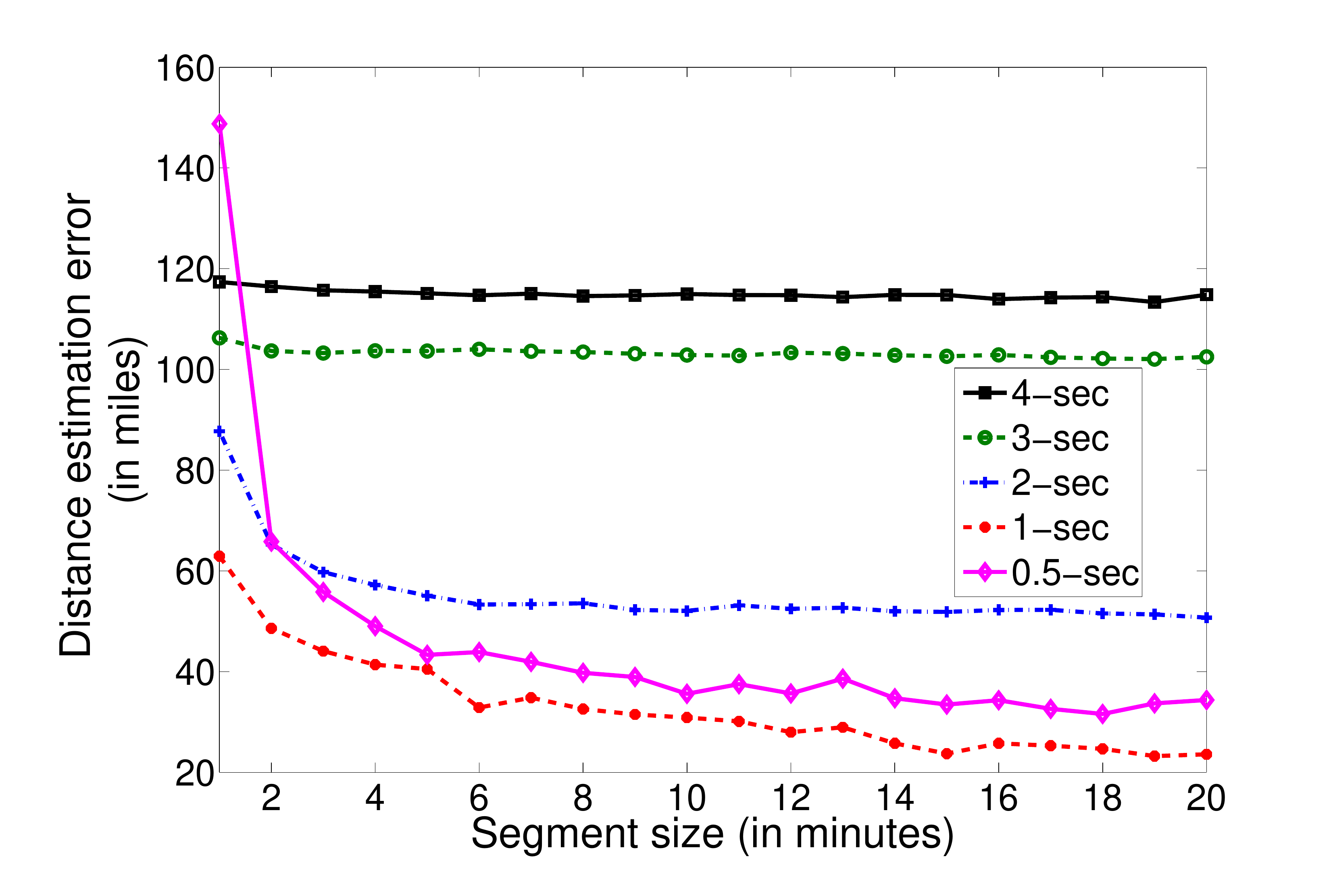}}
\end{center}
\caption{Mean error in distance estimation between Princeton and Atlanta using a linear relationship between correlation coefficients and geographical distances between the cities.}
\label{fig:AtlantaEstimateError}
\end{figure}

\paragraph{Distance estimation in Eastern US grid recordings} Let us denote Princeton,~NJ by City 1, College Park,~MD by City 2, and Atlanta,~GA by City 3. Assuming that city distances follows a linear relationship with the correlation coefficients, we use the values of correlation coefficients $\rho_{1,2}$, $\rho_{2,3}$, and corresponding city geographical distances $d_{1,2}$, $d_{2,3}$ to obtain an estimate of $d_{1,3}$ for a given observation of $\rho_{1,3}$. Based on the linear relationship, an estimate of $\widehat{d}_{1,3}$ for a given $\rho_{1,3}(n)$ can be given as:
\begin{equation}
\widehat{d}_{1,3}=d_{1,2}+ \frac{d_{1,2}-d_{2,3}}{\rho_{1,2}-\rho_{2,3}}(\rho_{1,3}-\rho_{1,2}).
\end{equation}

We compute the mean distance estimation error by averaging the absolute difference between the estimate $\widehat{d}_{x,y}(n)$ and the geographical distance $d_{x,y}$ for different query segments. The geographical distances between the cities are measured using Google maps by entering each city in the search and routing for directions using major highways. The plot of the mean distance error in distance estimation for different segment lengths and filter orders is shown in Figure~\ref{fig:AtlantaEstimateError}(a). According to this figure, when the filter of order $M=$~3 is used, a 15-minute long segment can provide a distance estimate within an accuracy of about 24 miles. Increasing the filter order degrades the distance estimates because the use of more data in filtering of the ENF signal averages the effects propagated due to the finite propagation speed of the frequency disturbances across the grid. To understand the effect of temporal resolution in distance estimation, we fix $M=$~3 and plot the average distance estimation error for different durations of instantaneous frequency (ENF) estimation in Figure~\ref{fig:AtlantaEstimateError}(b). From this figure, we observe that the best estimates are obtained when the instantaneous ENF is estimated every 1 second. Such a phenomenon can be explained by the finite speed of signal propagation, which is empirically determined to be on the order of $\approx$500 miles/second for the Eastern US grid~\cite{ENF_propagation}. As we increase the duration of data for instantaneous frequency estimation, the effect of the signal propagation averages out, leading to a decrease in the accuracy of distance estimates. Decreasing the signal duration for instantaneous frequency estimation to less than 1 second leads to an error in the estimated frequency itself due to the small number of data samples available for estimation.

\begin{figure}
\begin{center}
\includegraphics[width=0.4\textwidth]{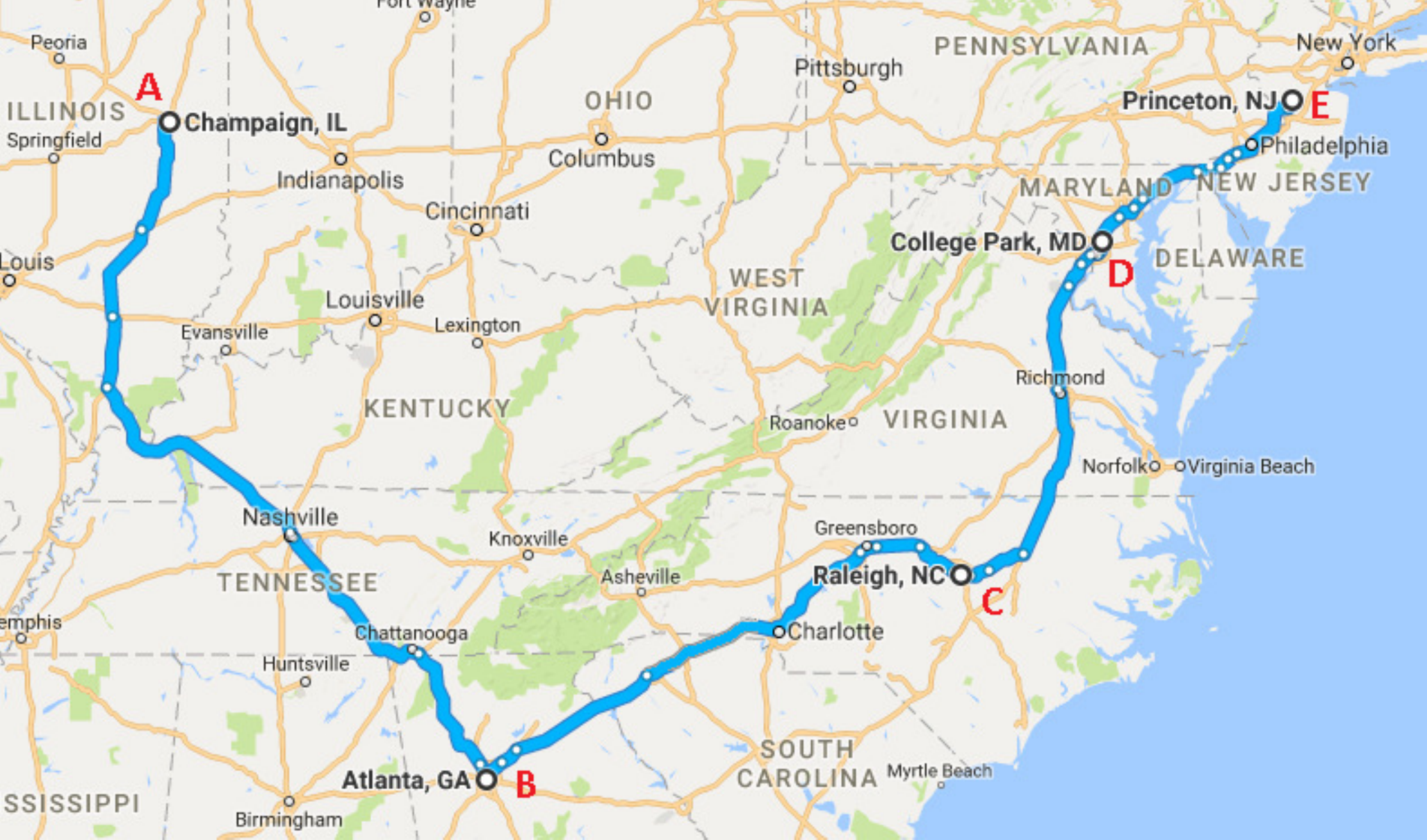}
\end{center}
\caption[Five locations shown on a map for Case Study 2]{Five locations shown on a map for Case Study 2: (A) Champaign, IL, (B) Raleigh, NC, (C) Atlanta, GA, (D) College Park, MD, and (E) Princeton, NJ. (Best viewed in color)}
\label{fig:5LocMap}
\end{figure}

\begin{figure*}
\begin{center}
\subfigure[Correlation with Princeton data]{\includegraphics[width=.32\textwidth]{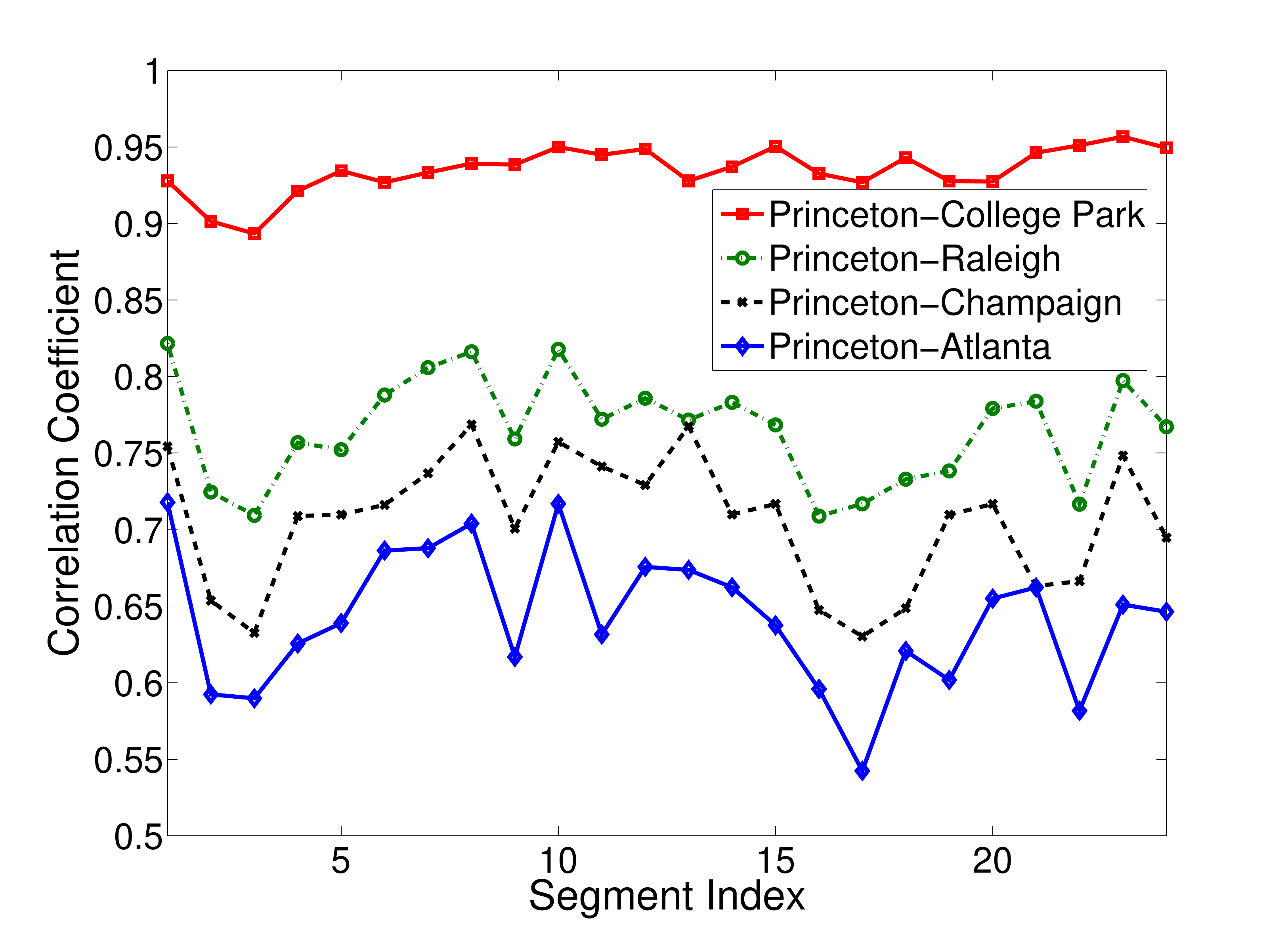}}
\subfigure[Correlation with College Park data]{\includegraphics[width=.32\textwidth]{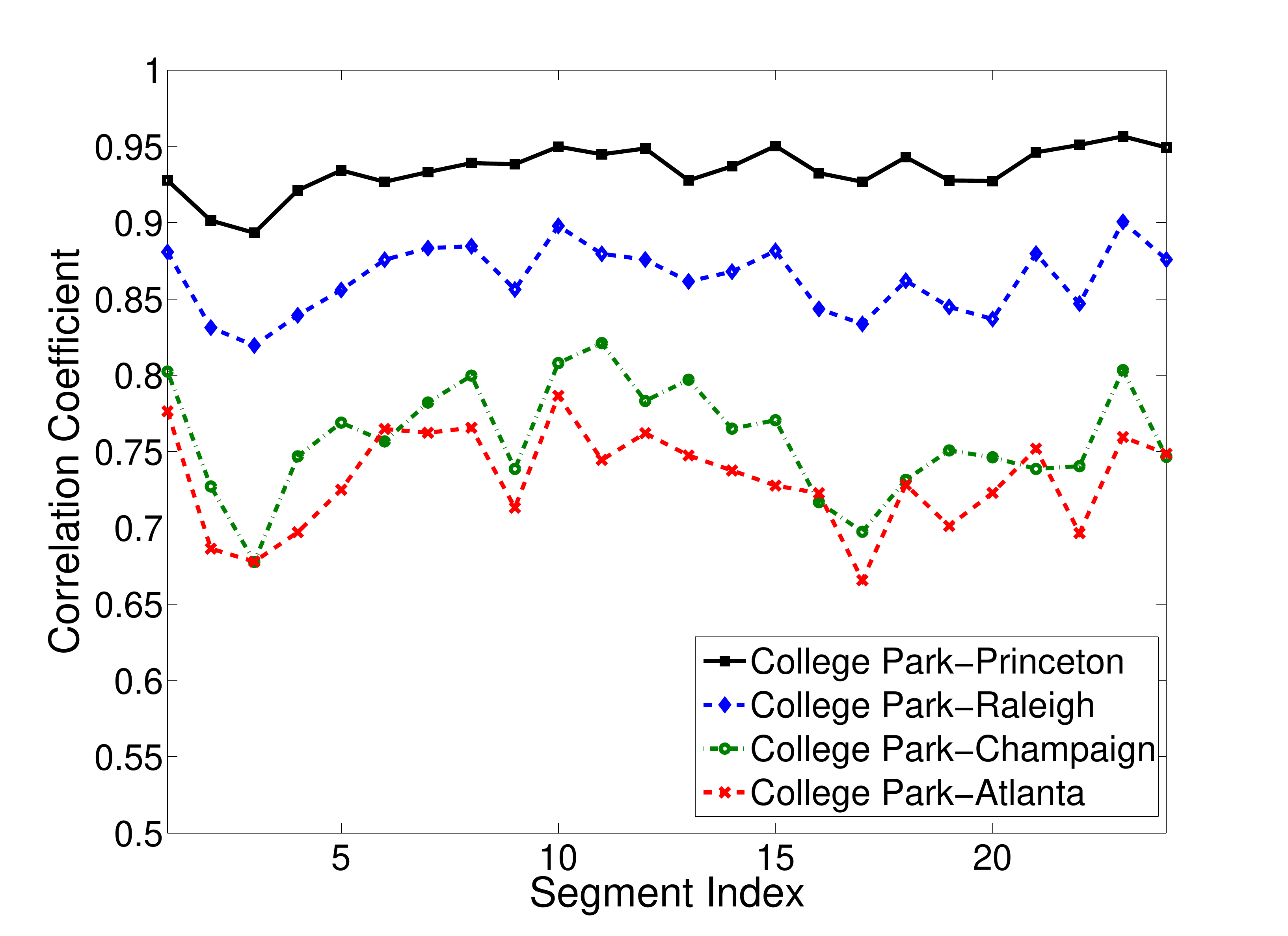}}
\subfigure[Correlation with Raleigh data]{\includegraphics[width=.32\textwidth]{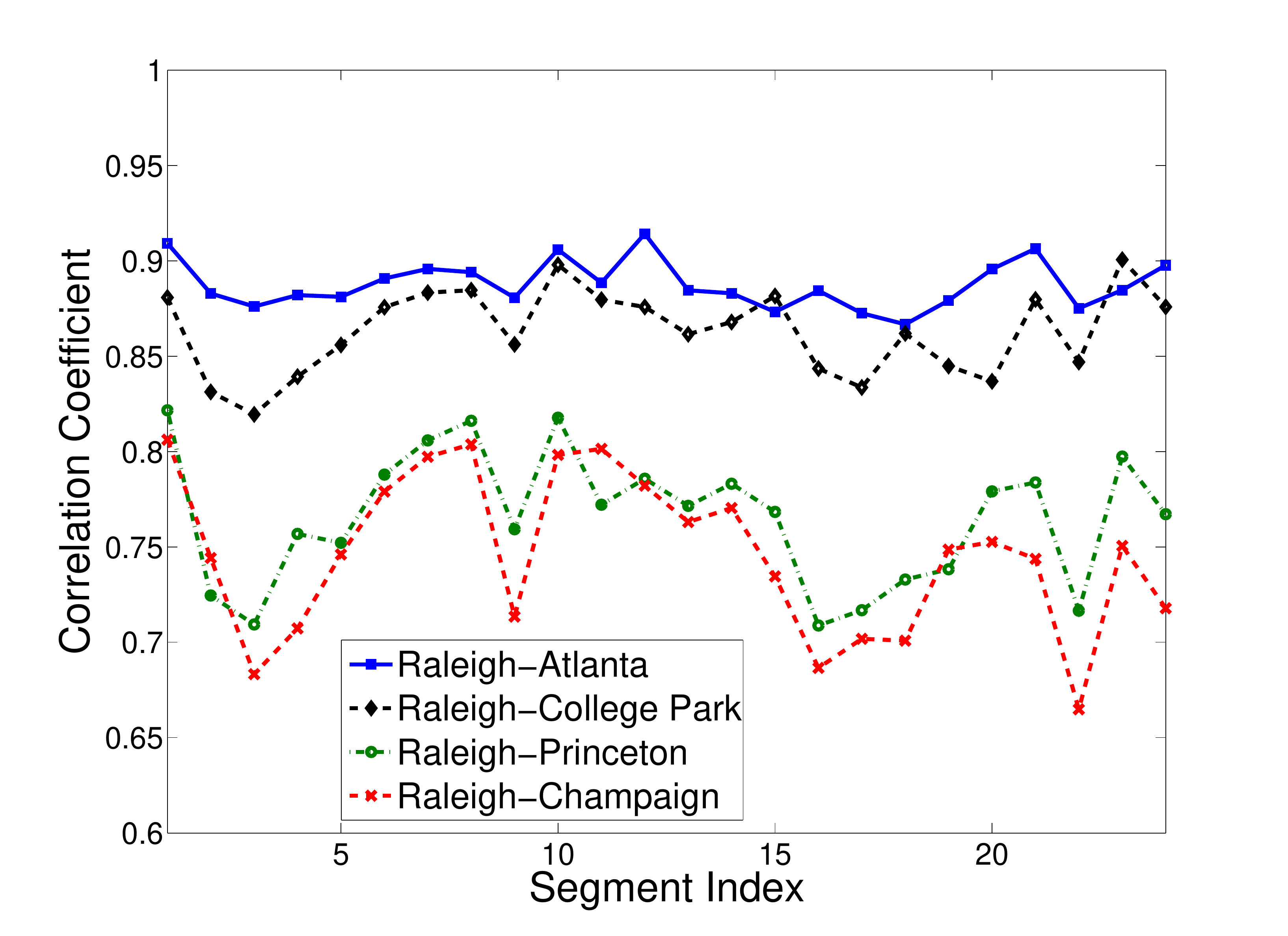}}
\subfigure[Correlation with Atlanta data]{\includegraphics[width=.32\textwidth]{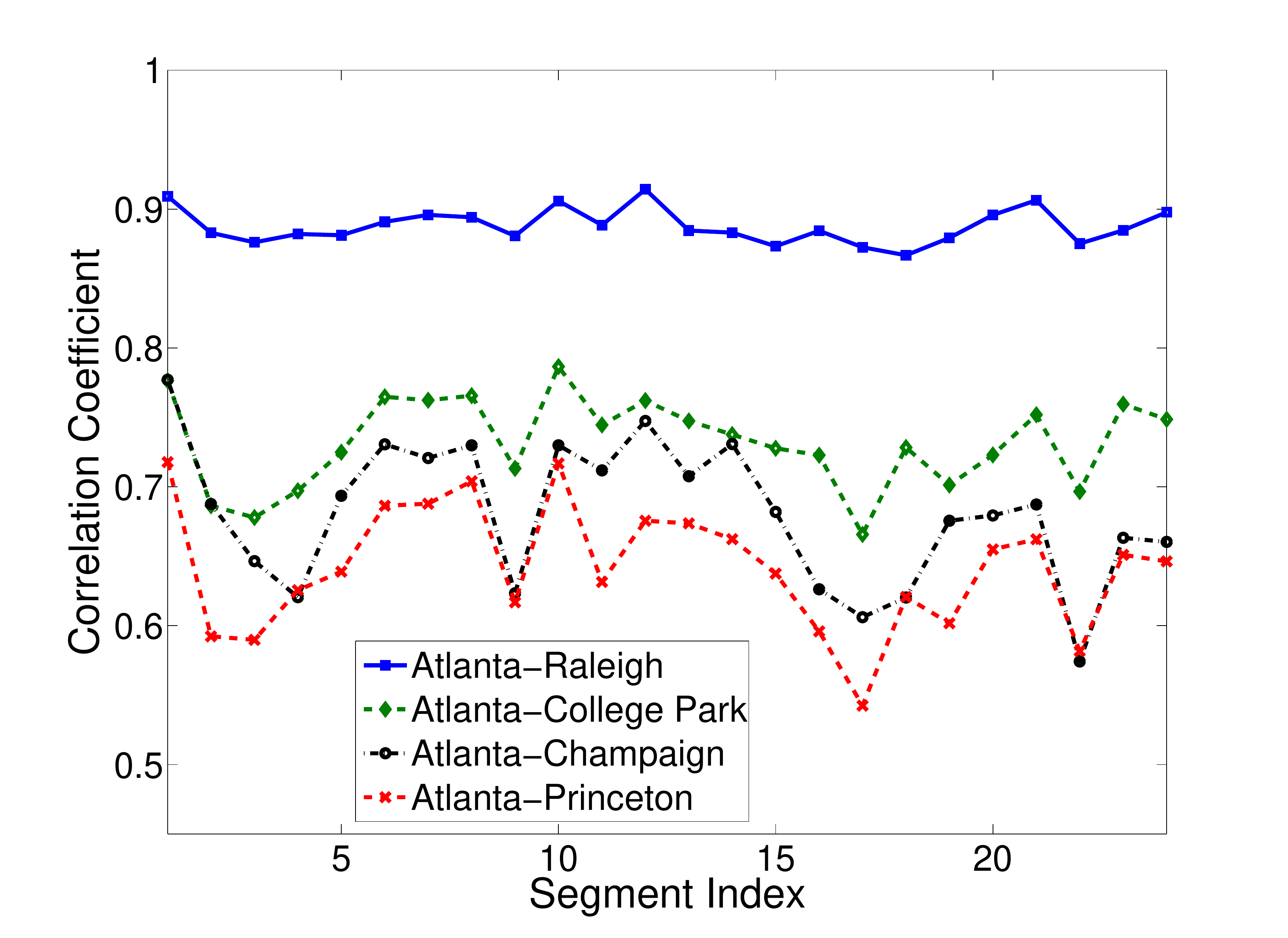}}
\subfigure[Correlation with Champaign data]{\includegraphics[width=.32\textwidth]{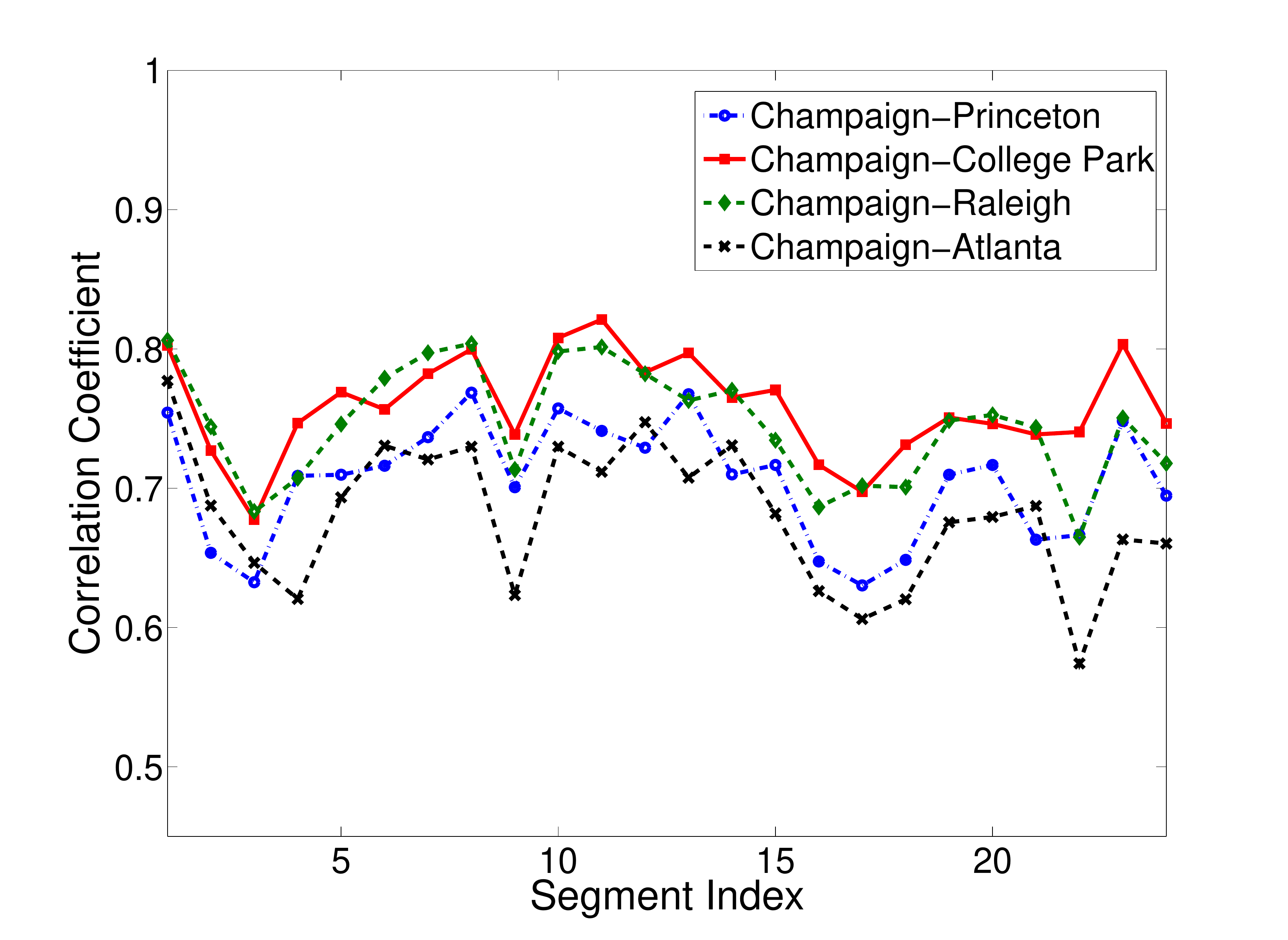}}
\end{center}
\caption[Correlation coefficient between the processed ENF signals for 5-location data in the Eastern US grid]{Correlation coefficient between the processed ENF signals across different locations for 10-minute long query segment for 5-location data in the Eastern US grid.}
\label{fig:5Location}
\end{figure*}

\paragraph{ENF as location-stamps} Based on this case study, we see that ENF signals have the potential to be used as location-stamps. The correlation coefficients of the signal extracted from a recording made at an unknown location with those extracted from data made at known locations can be used to estimate the distance of the recording location from the known locations. Known locations can behave as anchor nodes in designing localization protocols~\cite{Ravi_Localization}. Since the part of the grid used in this study is in densely populated region, geographical distance is a good approximation of the wireline distance. For sparsely populated areas in the grid and areas covering mountainous landscape, the wireline distance will be elongated as compared to geographical distance. The correlation coefficient-distance relationship in such sparsely populated regions of the grid may be further rectified by adding more locations to the anchor location dataset. In the next section, we discuss another case study on a 5-location dataset in the Eastern US grid that reveals additional challenges.

\subsection{Case Study 2: 5-Location Data from the Eastern US Grid}
\label{subsec:ch6:5LocationData}
For this experiment, power data was collected from two additional locations, Champaign in Illinois and Raleigh in North Carolina. The five locations are designated on a map in Figure~\ref{fig:5LocMap}. This 5-location data is four hours in duration. We use a similar sliding window based approach, as discussed in Sec.~\ref{subsec:ch6:ENFSignalProc}, to temporally align the signals, and estimate the correlation coefficients between the data from different city pairs after filtering through a high pass filter of order $M=$~3 for each segment length of 10~minutes. The plots of the correlation coefficients between data from different cities are shown in Figure~\ref{fig:5Location}. From these figures, we observe that the correlation coefficients between the data collected from cities closer to each other are higher than those from cities further apart, similar to the observation from 3-location data. The relative magnitude of the correlation coefficients are roughly inversely proportional with the geographical distance between the cities. For example, the distance between College Park and Princeton is the least mutual distance among all city pairs, and the correlation coefficient between the data collected from these two locations is the highest. However, due to a different grid density and 2-dimensional relations of the electricity flows, it may not be possible to use the straight line assumption that was used in Section~\ref{subsec:ch6:3Locations}.

Figure~\ref{fig:GridDensityUS} shows a grid density map of the Eastern US interconnection grid from which we observe that the grid-density is non-uniform at different places and along different directions. As the flow of ENF variations over the wire lines depends on such parameters as grid topology (road distance may not be the same as the actual wire distance), grid density, etc., the correlation coefficient between data from different locations may have a complex relationship with the geographical distance between these locations. Limited amount of data is available to us, so we must design a localization protocol without learning an explicit relationship between the correlation coefficient and the distance between different locations. Instead, we use the observation from our experiments that the pair-wise correlation between the locations far apart is less than the pair-wise correlation between the closer cities. Using such observations, we devise a method of half-plane intersection to estimate an unknown location of recording.

\section{Methods for Localization}
\label{sec:ch6:methods}
In this section, we explore two methods for ENF-based fine localization of power recordings within a grid: the half-plane intersection method and the correlation quantization method. We also examine if the results can be improved by a combination of both methods.

\subsection{Half-Plane Intersection for Localization}
\label{subsec:ch6:halfplan}

Let us denote the location of $K$ cities by $P_1=\{x_1, y_1\}, P_2=\{x_2, y_2\}, \ldots, P_K=\{x_K, y_K\}$. Suppose we are given ENF data collected at all anchor cities, along with their known locations. Based on this information, we derive a localization protocol to estimate the unknown location of a city (denoted by $P_{Query}$), based on the ENF data recorded at that location. We assume that the query city location $P_{Query}$ and the locations of all anchor nodes lie in a rectangular region surrounding $P_1, P_2, \ldots, P_K$ and denoted by $D$. We refer to $D$ as the domain of the localization region. As discussed in Section~\ref{subsec:ch6:5LocationData}, if the distance between $P_i$ and $P_{Query}$ is greater than the distance between $P_j$ from $P_{Query}$, we generally have $\rho_{j,Query}>\rho_{i,Query}$. Based on this observation, we claim that the estimated $\widehat{P}_{Query}$ lies in the half-plane given by the region denoted by the set of locations $\widehat{P}_{i,j}$:

\begin{small}
\begin{equation}
\nonumber
X  : ||X-P_i||_2 > ||X-P_j||_2, X \in D, \mathrm{if } \rho_{j,Query}-\rho_{i,Query}>0,
\end{equation}
\begin{equation}
\label{eq:halfplane}
X  : ||X-P_i||_2 \leq ||X-P_j||_2, X \in D, \mathrm{if } \rho_{j,Query}-\rho_{i,Query} \leq 0.
\end{equation}
\end{small}

The conditions described in Eq.~\eqref{eq:halfplane} are the sign bit of the difference between the correlation coefficients, and they use highly quantized information from the correlation coefficient. The conditions also provide us with hard decision boundaries for the half-plane and do not take into account the noisy nature of pair-wise correlation coefficients. For example, when the correlation coefficients of the query city's ENF signal with $i^{th}$ and $j^{th}$ locations are close to each other, i.e., if $|\rho_{i,Query}-\rho_{j,Query}| < \epsilon$ for a small $\epsilon$, the confidence in assigning a half-plane to the feasible solution set $\widehat{P}_{i,j}$ is reduced in Eq.~(\ref{eq:halfplane}). To compensate for such values of correlation coefficients, we replace the feasible set given by Eq.~(\ref{eq:halfplane}) with the following equation with a tolerance $\epsilon$, where $\widehat{P}_{i,j}$ would be equal to:

\begin{small}
\begin{equation}
\nonumber
X: ||X-P_i||_2 > ||X-P_j||_2, X \in D,  \text{if } \rho_{j,Query}-\rho_{i,Query} \geq \epsilon,
\end{equation}
\begin{equation}
\label{eq:halfplaneNew}
X: ||X-P_i||_2 \leq ||X-P_j||_2, X \in D,  \text{if } \rho_{j,Query}-\rho_{i,Query} \leq \epsilon.
\end{equation}
\end{small}
Using the correlation value obtained from all the anchor nodes, the set of feasible points can be reduced further by computing the intersection of all the feasible half-planes as follows:
\begin{equation}
\label{eq:halplaneintersection}
\widehat{P}_{Query}=\cap_{i,j}\widehat{P}_{i,j} \hspace{4mm} i,j\in\{1,2,\ldots ,K\}, i\neq j.
\end{equation}

\begin{figure*}
\begin{center}
\subfigure[Constraint 1 (College Park-Raleigh)]{\includegraphics[width=.32\textwidth]{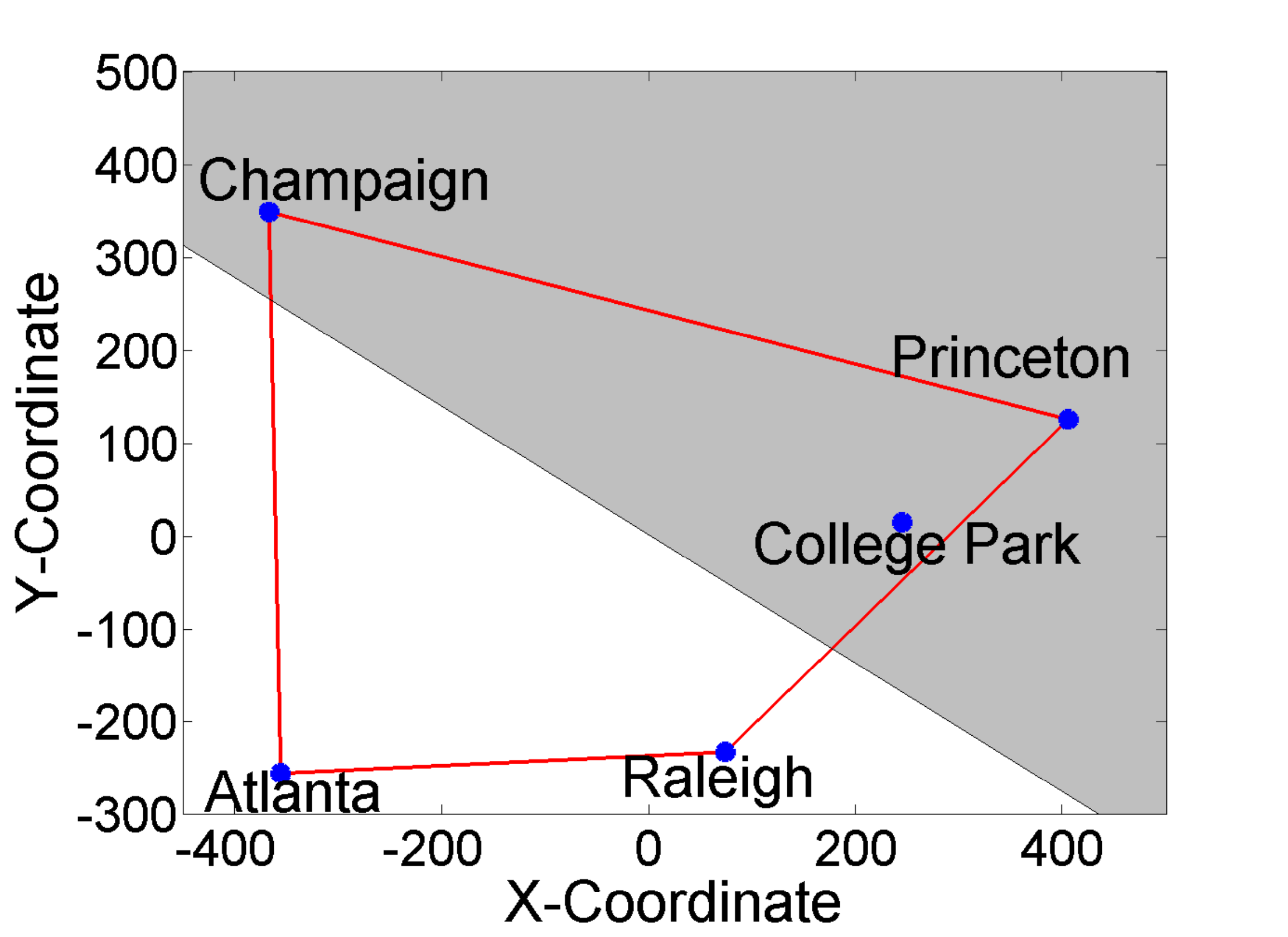}}
\subfigure[Constraint 2 (College Park-Atlanta)]{\includegraphics[width=.32\textwidth]{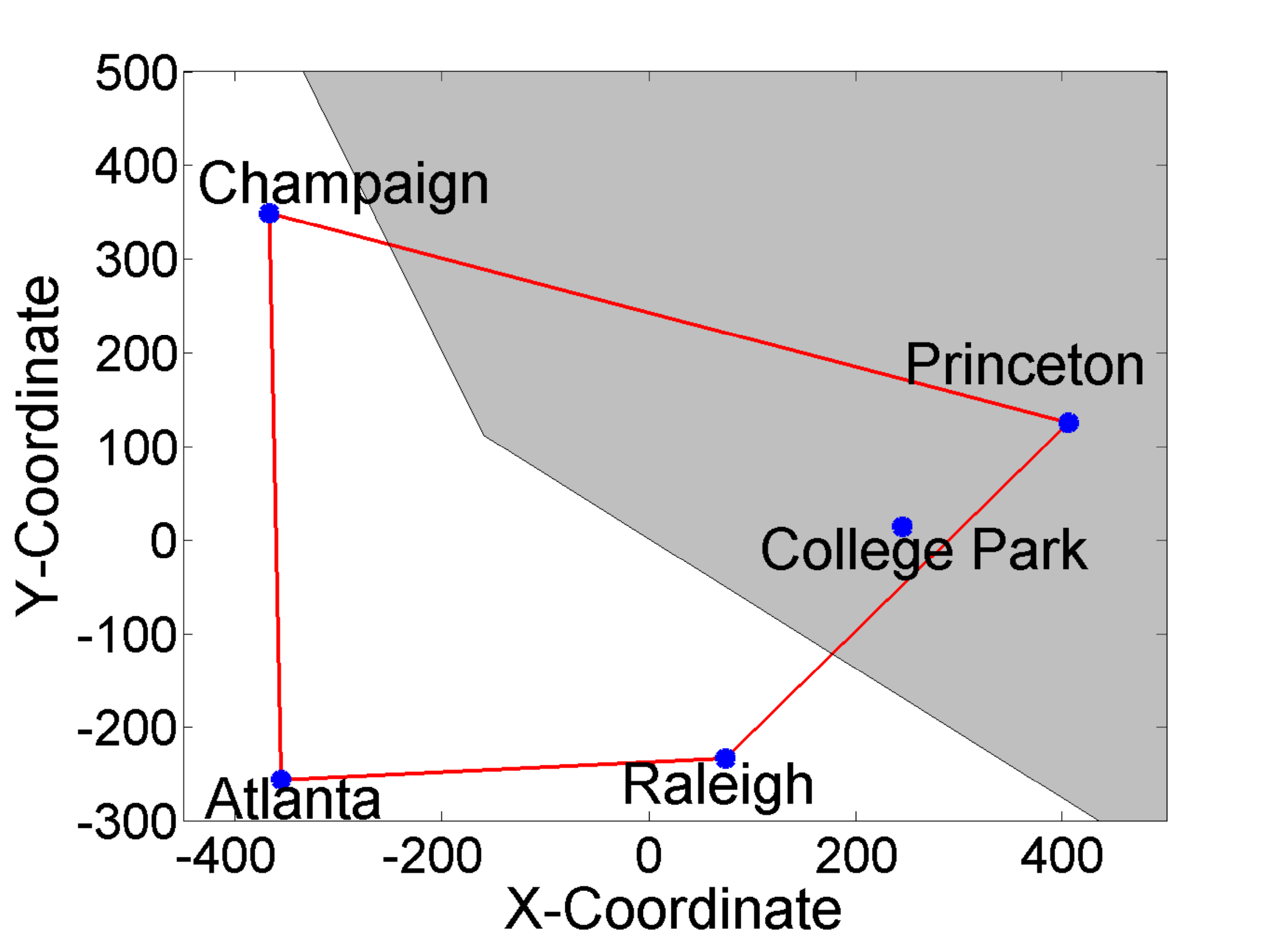}}
\subfigure[Constraint 3 (College Park-Champaign)]{\includegraphics[width=.32\textwidth]{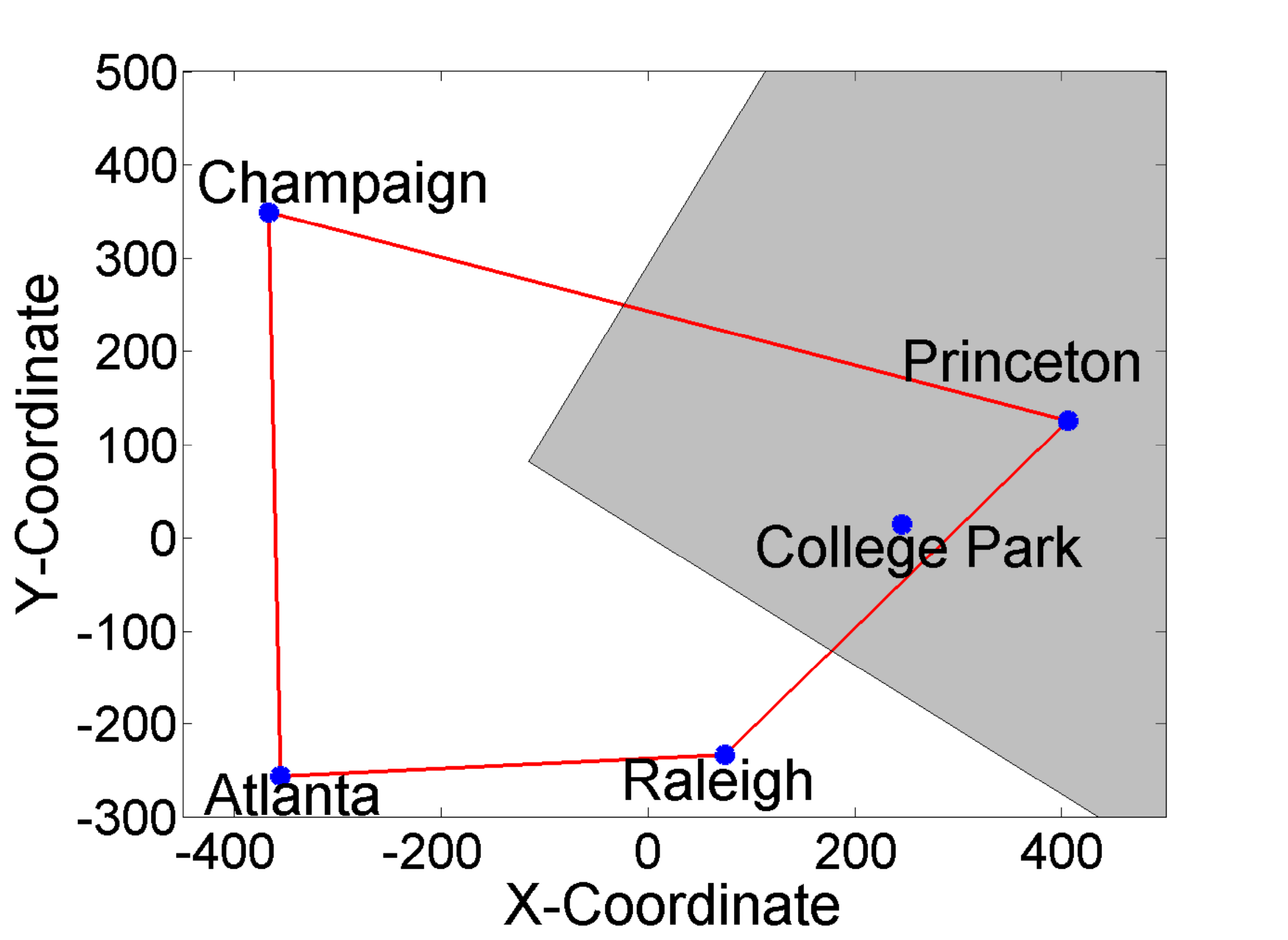}}
\subfigure[Constraint 4 (Raleigh-Atlanta)]{\includegraphics[width=.32\textwidth]{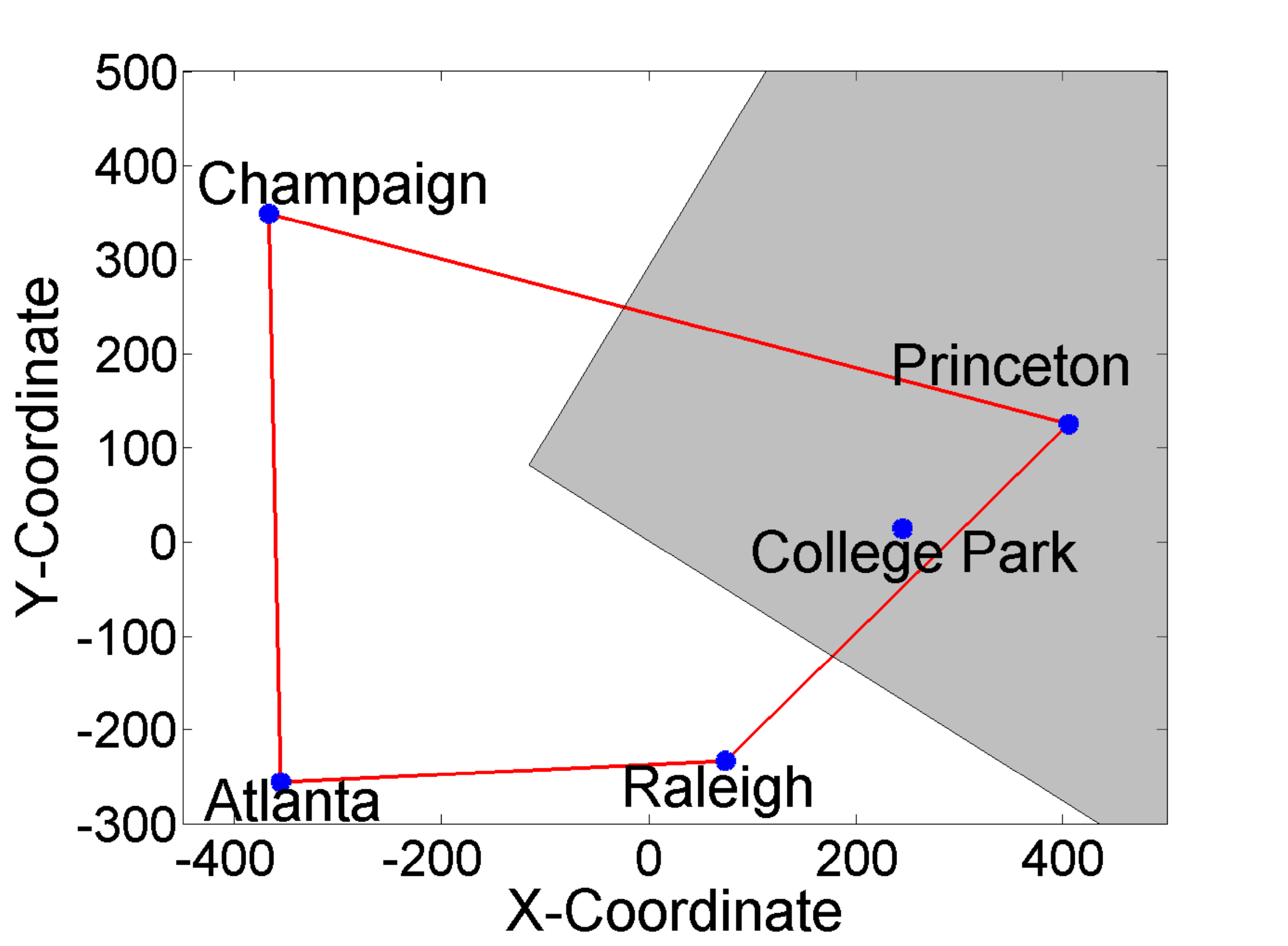}}
\subfigure[Constraint 5 (Raleigh-Champaign)]{\includegraphics[width=.32\textwidth]{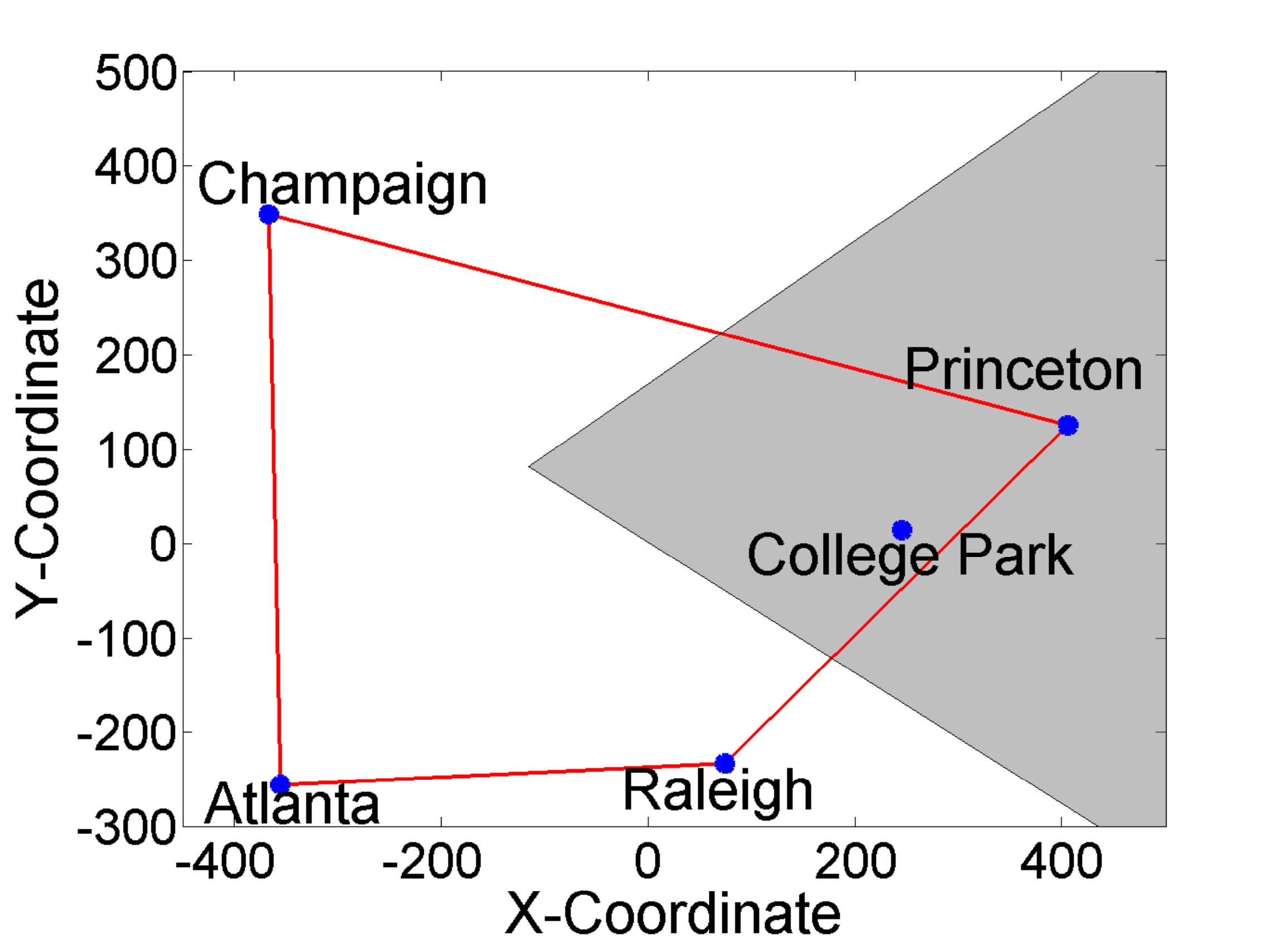}}
\subfigure[Constraint 6 (Champaign-Atlanta)]{\includegraphics[width=.32\textwidth]{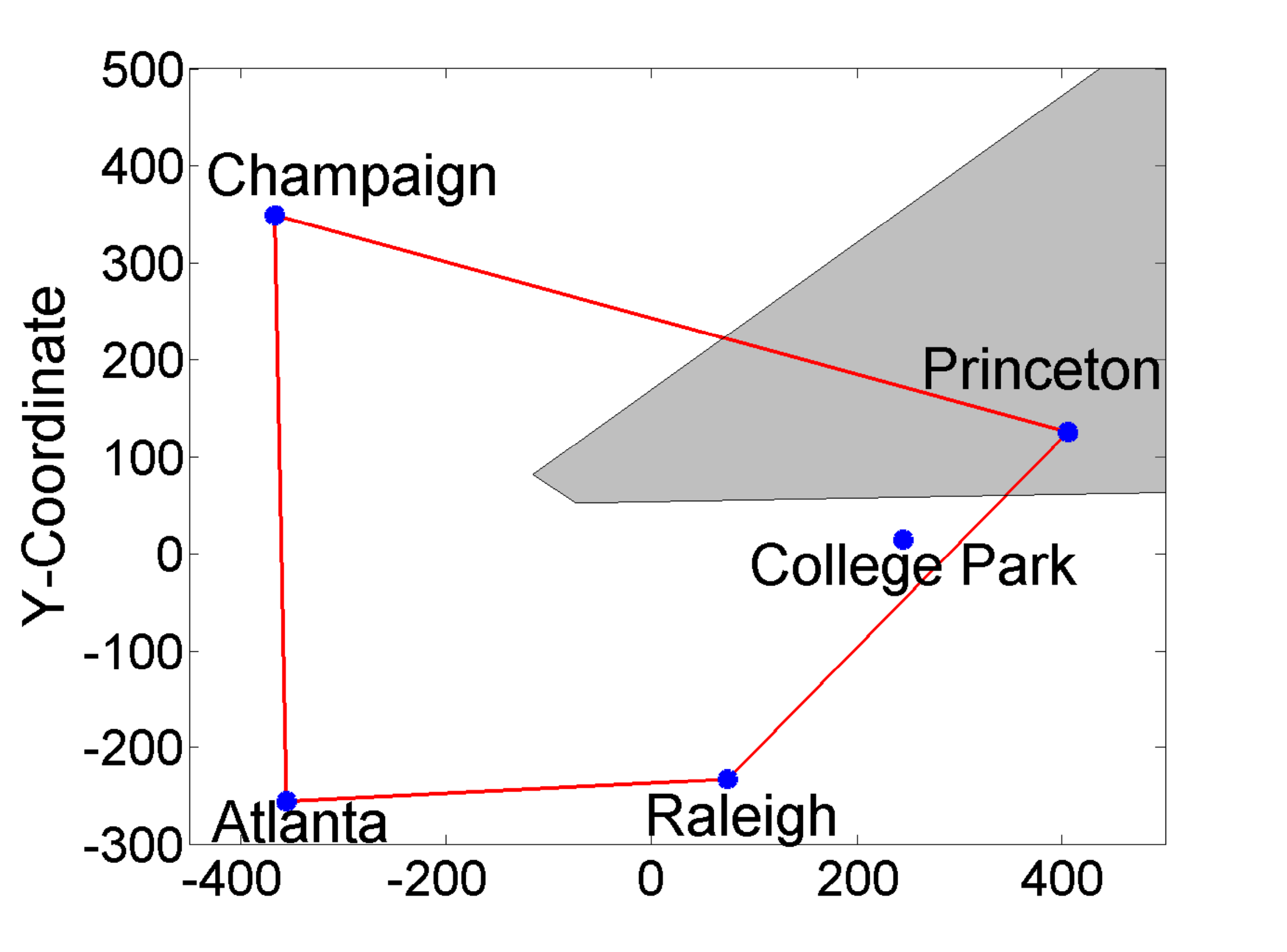}}
\end{center}
\caption[Localization example for Princeton using half-plane intersection]{Example of localization of Princeton using the half-plane intersection method, when College Park, Raleigh, Atlanta, and Champaign are the anchor nodes. Each sub-figure caption pair represents the anchor cities recordings used for deriving half-plane constraints. Shaded area represents the feasible region after each additional distance-correlation constraint defined in  Eq.~\eqref{eq:halfplane} is applied. The relative positions of the cities with respect to each other are shown.}
\label{fig:Example5Loc}
\end{figure*}

The order of constraints does not have any impact on the estimated feasible area because intersections of half-planes, derived from each constraint, are used to estimate the resulting feasible area. As we have ENF data from five locations, we use four locations as anchor cities and use the ENF data of the fifth city as the query data to estimate its location using the proposed half-plane intersection method. In Figures~\ref{fig:Example5Loc}(a)--\ref{fig:Example5Loc}(f), we show an example of the set of feasible regions obtained after adding each half-plane constraint, when College Park, Raleigh, Atlanta, and Champaign are the anchor cities and Princeton is the query city. In these figures, the shaded region represents the estimated feasible region for the query city. From this figure, we observe that the area of the feasible region decreases with an increase in the number of constraints, meaning that the precision of localization improves.

Due to the limited amount of data and huge geographical area encompassed by the anchor node locations in our experiments, we define the localization performance of the proposed method in terms of two metrics: the probability of localization, denoted by $p_{loc}$, and the area of localization, denoted by $a_{loc}$. If data from more anchor cities are available, location estimates can be defined using such a metric as the centroid of the feasible set. The first performance metric, $p_{loc}$, measures the fraction of queries for which the feasible region contains the true location of the query city, i.e.,
\begin{equation}
\label{eq:PLoc}
p_{loc}=\frac{\mbox{\# of queries for which } \widehat{P}_{Query} \mbox{ contains $P_{Query}$}}{\mbox{\# of queries}},
\end{equation}
and determines the performance of the proposed method in terms of classifying the region of the query city. A higher value of $p_{loc}$ indicates that the proposed method can assign the query city to a correct feasible region with a high probability. The second performance metric, $a_{loc}$, measures the ratio of the area of the feasible set with respect to the area of the total domain on which the localization is performed, for cases when the feasible set contains the location of the query city. The domain can be defined, for example, as a rectangular region that surrounds all the five locations in our dataset. $a_{loc}$ can be mathematically represented as:
\begin{equation}
\label{eq:ALoc}
a_{loc}=\frac{\mbox{area of region } \widehat{P}_{Query}}{\mbox{area of domain } D}.\mathbbm{1}(P_{Query}\in \widehat{P}_{Query}),
\end{equation}
where $\mathbbm{1}(\cdot)$ is an indicator function. $a_{loc}$ determines the precision of the localization, i.e., the smaller the value of $a_{loc}$, the higher is the localization precision.

\begin{figure}
\begin{center}
\subfigure[4-minute query segment]{\includegraphics[width=.32\textwidth]{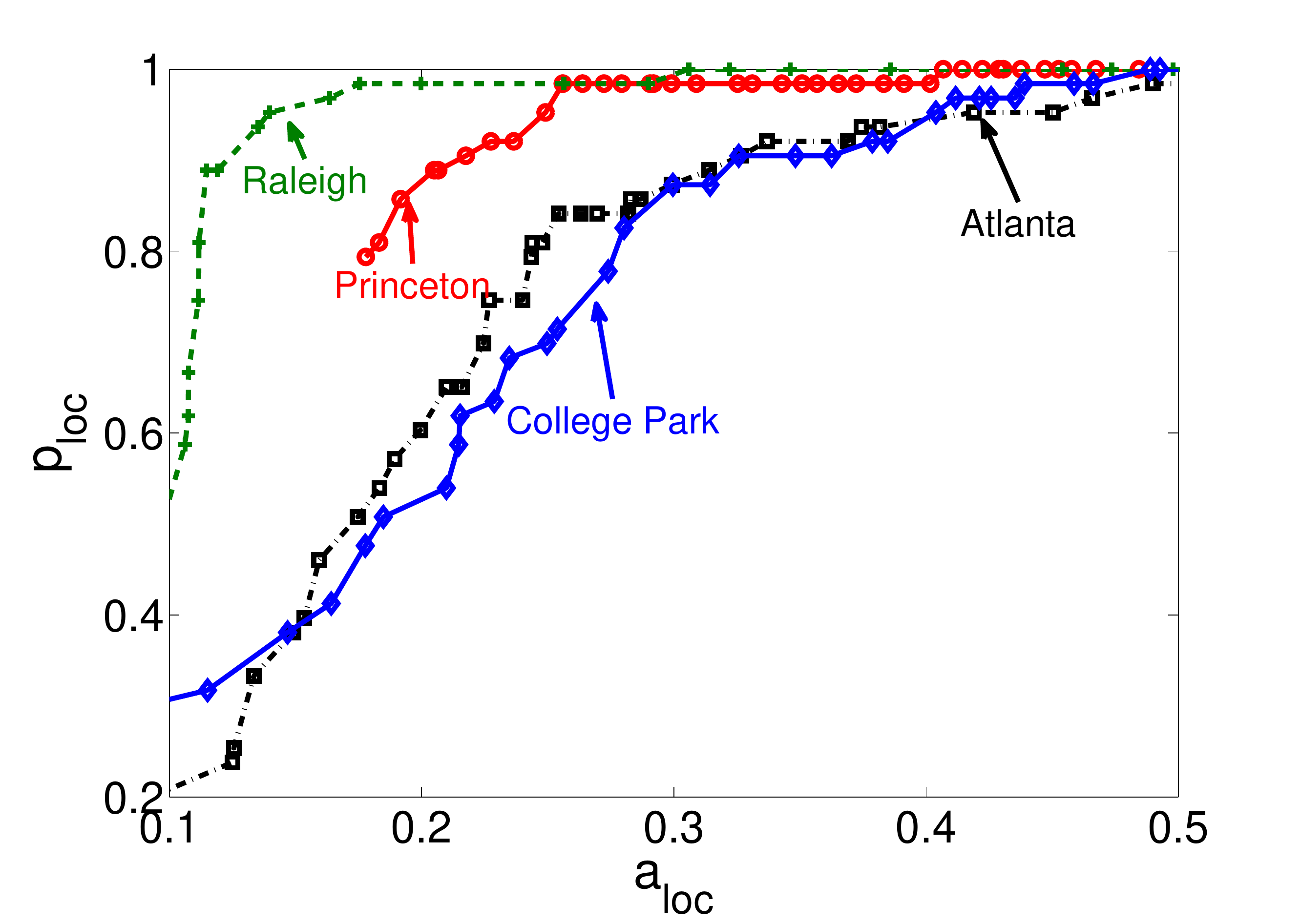}}
\subfigure[8-minute query segment]{\includegraphics[width=.32\textwidth]{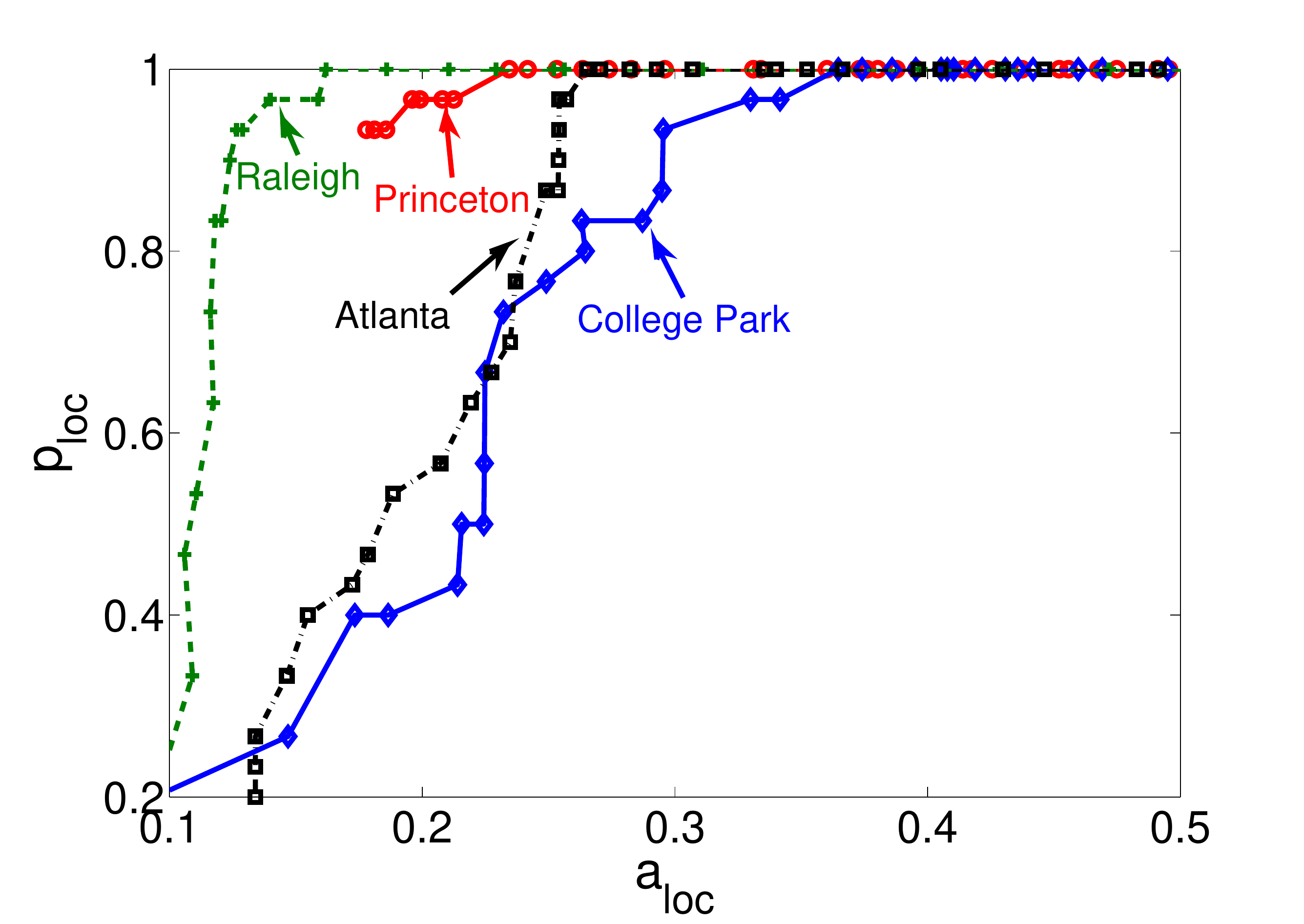}}
\end{center}
\caption{Probability of localization, $p_{loc}$, and area of localization, $a_{loc}$ for 5-location Eastern US data using the half-plane intersection method.}
\label{fig:PLoc}
\end{figure}

Figures~\ref{fig:PLoc}(a)--(b) show the plots of $a_{loc}$ vs. $p_{loc}$ for different cities by considering the other four cities in the dataset as anchor nodes for four-minute long and eight-minute long query segments, respectively. The different values of $a_{loc}$ and $p_{loc}$ are obtained by varying the value of $\epsilon$. In an ideal scenario, the value of $p_{loc}$ should be close to one and the value of $a_{loc}$ should be close to zero for an accurate localization. From this plot, we observe that $p_{loc}$ increases with an increase in the value of $a_{loc}$ for all the cities (equivalently, the precision decreases). This happens due to the trade-off introduced by the tolerance $\epsilon$ in $p_{loc}$ and $a_{loc}$. For low values of $\epsilon$, the value of $p_{loc}$ is less, as the hard decision rule does not provide a correct estimate of the feasible area of location of the query city when measurements are noisy. As the value of $\epsilon$ increases, hard decision boundaries provide a tolerance, which increases the value of $p_{loc}$, but a decrease in the number of constraints also reduces the precision. As the query segment duration increases from four to eight minutes, the localization performance also improves as using a large segment size provides a more robust estimate of the correlation coefficient.

\begin{figure}
\begin{center}
\subfigure[$\rho_{i,j}$ vs $d_{i,j}$]{\includegraphics[width=.32\textwidth]{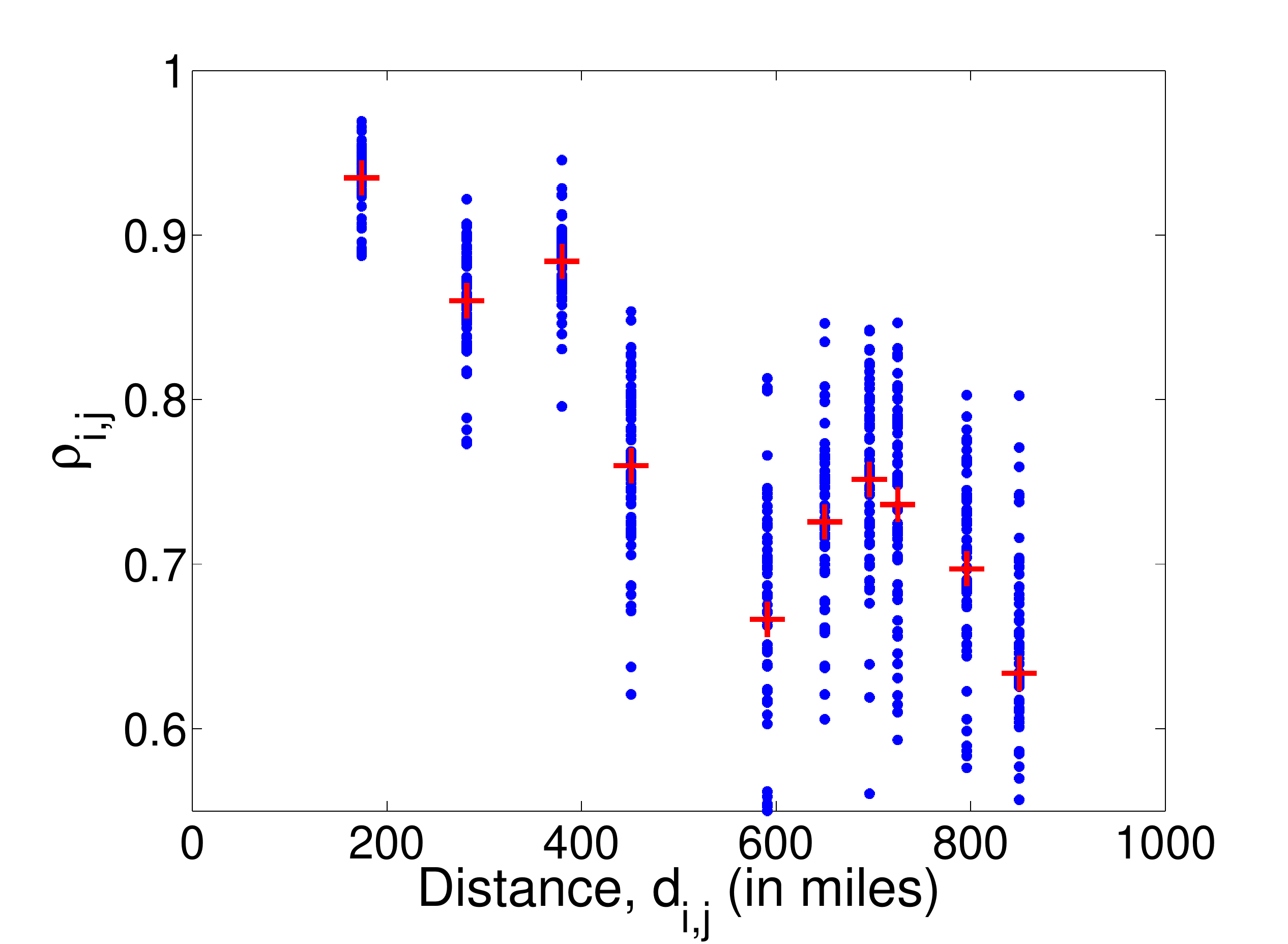}}
\subfigure[Quantization of $\rho_{i,j}$ and corresponding distance bins]{\includegraphics[width=.32\textwidth]{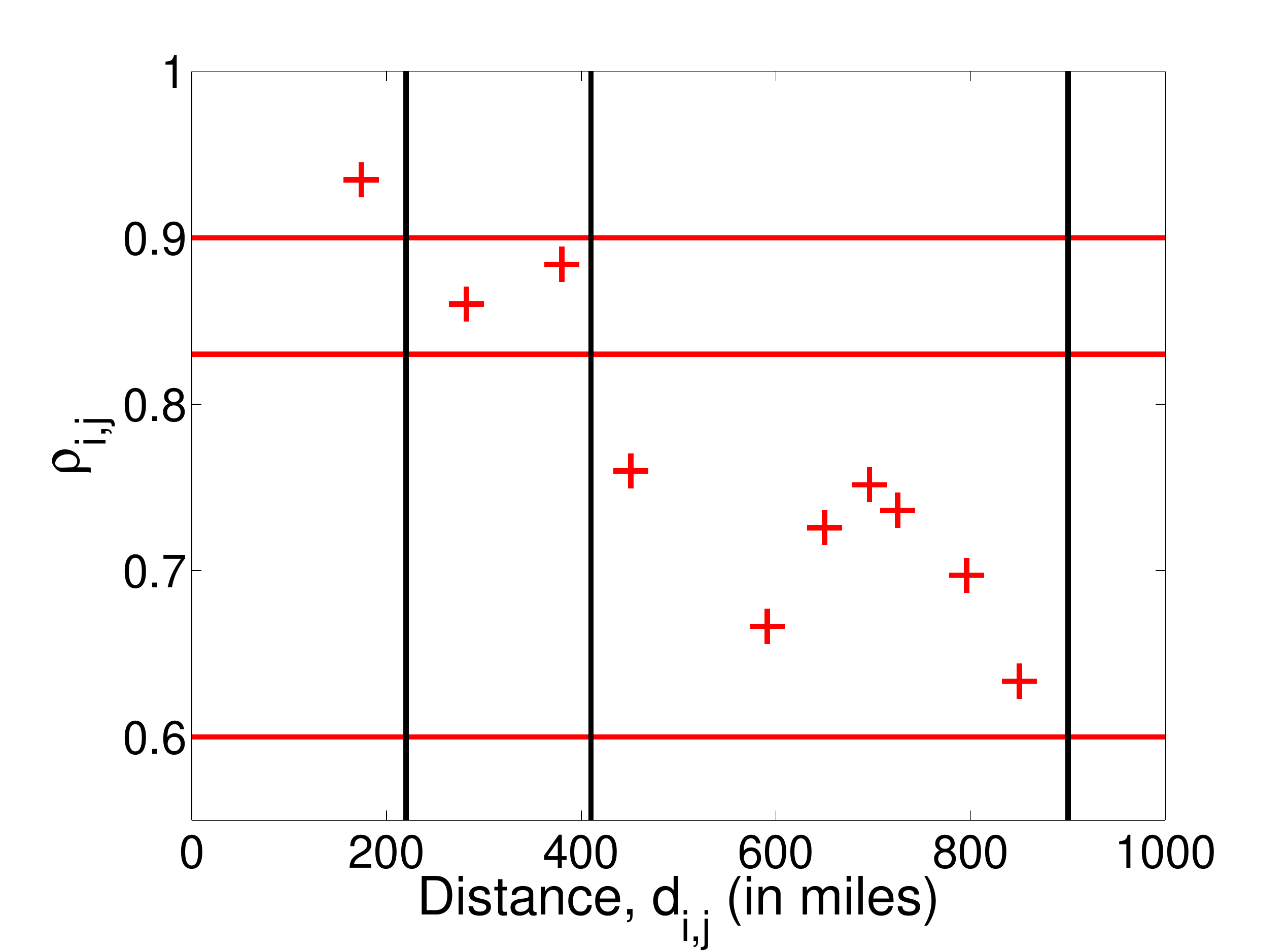}}
\end{center}
\caption{Relationship between $\rho_{i,j}$ and $d_{i,j}$ for 5-location data.}
\label{fig:RhovsD}
\end{figure}

\subsection{Correlation Quantization for Localization}
\label{subsec:ch6:halfplane}

\begin{figure}
\begin{center}
\subfigure[Constraint 1 (College Park)]{\includegraphics[width=.24\textwidth]{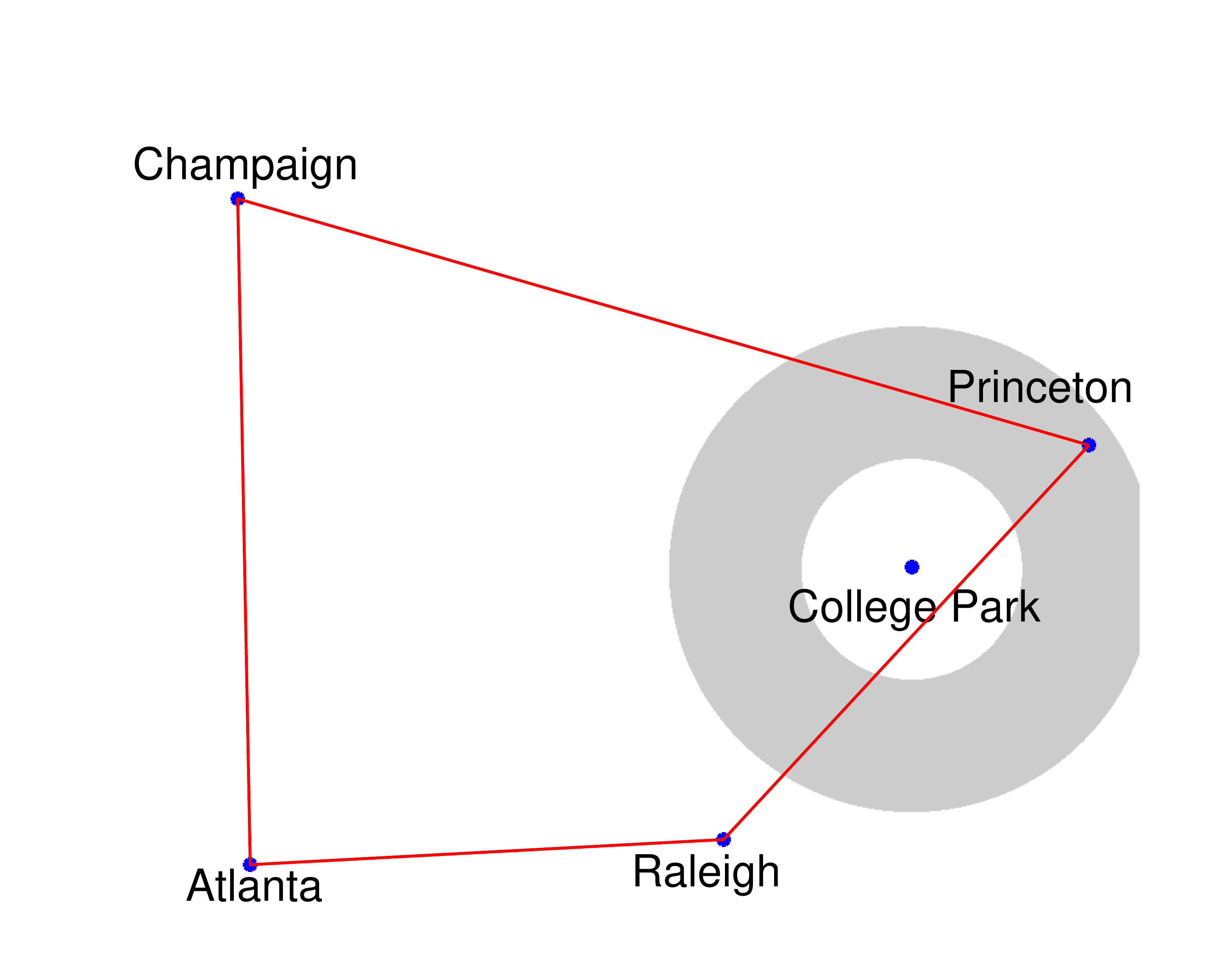}}
\subfigure[Constraint 2 (Raleigh)]{\includegraphics[width=.24\textwidth]{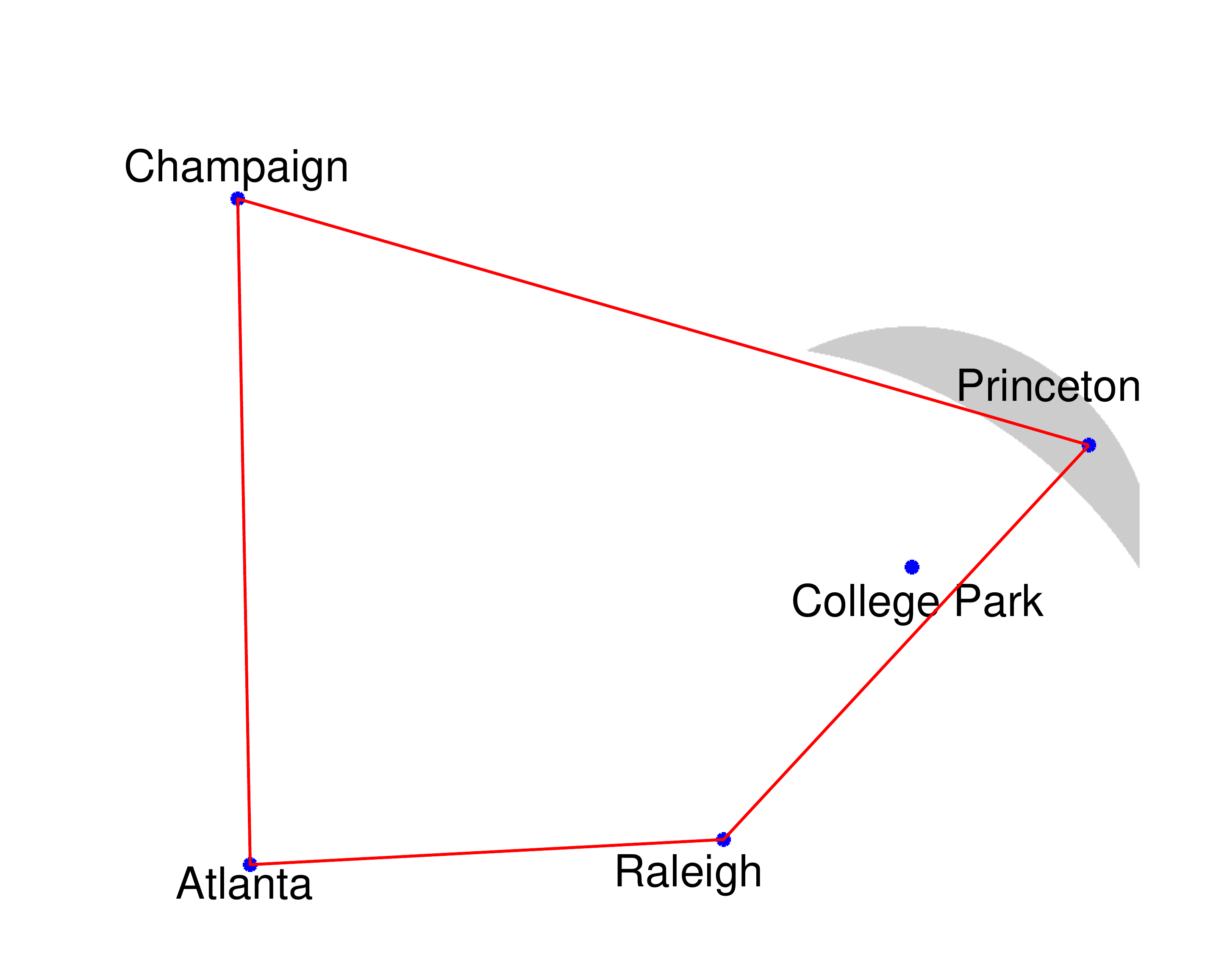}}
\subfigure[Constraint 3 (Atlanta)]{\includegraphics[width=.24\textwidth]{CircleExample5LocationPrincetonLocalization-eps-converted-to.pdf}}
\subfigure[Constraint 4 (Champaign)]{\includegraphics[width=.24\textwidth]{CircleExample5LocationPrincetonLocalization-eps-converted-to.pdf}}
\end{center}
\caption{Example of localization of Princeton using the correlation quantization method with College Park, Raleigh, Atlanta, and Champaign as anchor locations. Each sub-figure caption represents the anchor city recordings used for deriving circle constraints. Shaded area represents the feasible region after each additional distance-correlation constraint in  Eq.~\eqref{eq:AreaCorrelation} is applied.}
\label{fig:Example5LocCircle}
\end{figure}

\begin{figure}
\begin{center}
\subfigure[Half-plane method]{\includegraphics[width=.24\textwidth]{PrincetonLocalizationConstraint6-eps-converted-to.pdf}}
\subfigure[Correlation quantization method]{\includegraphics[width=.24\textwidth]{CircleExample5LocationPrincetonLocalization-eps-converted-to.pdf}}
\subfigure[Half-plane and correlation quantization method]{\includegraphics[width=.24\textwidth]{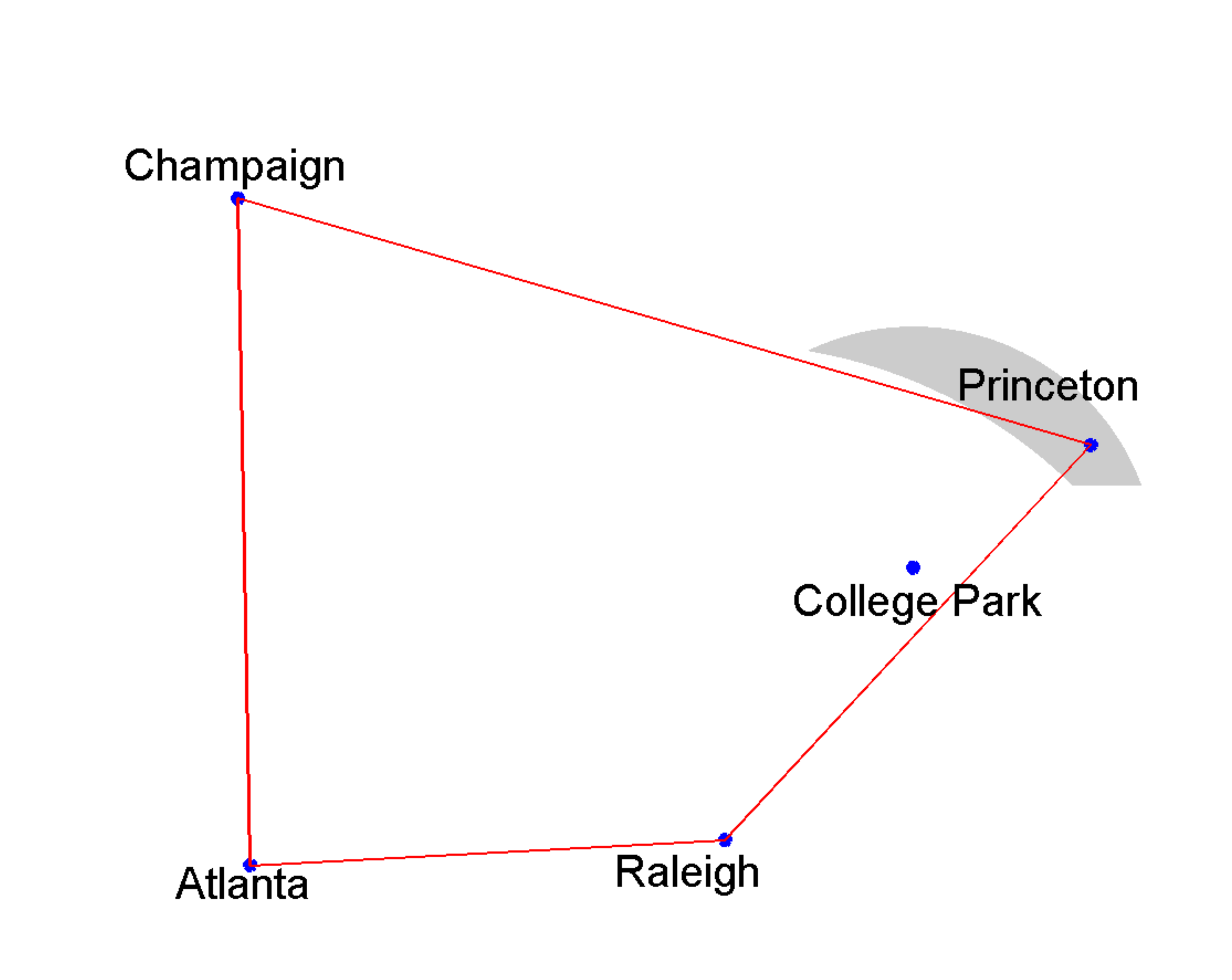}}
\end{center}
\caption{Example of localization of Princeton using all the three localization protocols with College Park, Raleigh, Atlanta, and Champaign as anchor locations. Shaded area represents the estimated feasible region of recording.}
\label{fig:ExampleAll}
\end{figure}

In this section, we describe a method similar to trilateration~\cite{Garg_ICASSP_Localization} for localization that makes use of the quantized correlation coefficient information between the city pairs. In this method, we rely on the observation from Figures~\ref{fig:5Location}(a)--\ref{fig:5Location}(d) that the correlation coefficient value $\rho_{i,j}$ between the location signatures of the $i^{th}$ and the $j^{th}$ location decrease as the distance $d_{ij}$ between them increases. We plot $\rho_{i,j}$ as a function of $d_{ij}$ for the 5-location data in Figure~\ref{fig:RhovsD}(a), which shows that the locations close to each other have a higher value of $\rho_{i,j}$ as compared to locations further apart. However, it is difficult to derive a functional relationship between the value of  $\rho_{i,j}$ and $d_{i,j}$. A close observation of Figure~\ref{fig:RhovsD}(a) reveals that it may be possible to quantize the values of $\rho_{i,j}$ in distance ranges. One realization of such a quantization is shown in Figure~\ref{fig:RhovsD}(b). In this figure, the red points indicate the mean of the value of $\rho_{i,j}$ at a particular distance, the red line indicates the quantization bin on $\rho_{i,j}$ axis, and the black lines indicate the corresponding distance bins. Based on such a quantization scheme, the following relationship can be derived between $\rho_{i,j}$ and $d_{i,j}$ for our 5-location dataset:

\begin{small}
\begin{eqnarray}
\label{eq:RhoDRelation}
\nonumber 100 \leq d_{i,j} < 220, & \text{if } 0.9<\rho_{i,j} \leq 1 \\
\nonumber 220 \leq d_{i,j} < 450, & \text{if } 0.83<\rho_{i,j} \leq 0.9 \\
\nonumber 450 \leq d_{i,j} < 900, & \text{if } 0.55<\rho_{i,j} \leq 0.83 \\
900 \leq d_{i,j},  & \text{if } <\rho_{i,j} \leq 0.55.
\end{eqnarray}
\end{small}

Based on the relations in Eq.~\eqref{eq:RhoDRelation}, a trilateration based protocol can be derived to estimate the feasible region of recording for a query city. In this trilateration protocol, using the constraints from Eq.~(\ref{eq:RhoDRelation}), the following relations are derived between the correlation coefficient and the distance between a query location and an anchor location, to estimate $\widehat{P}_{i,Query}$:

\begin{small}
\begin{eqnarray}
\label{eq:AreaCorrelation}
\nonumber X: 100 \leq \parallel X-P_i \parallel_2 < 220, X \in D,& 0.9<\rho_{i,Query} \leq 1 \\
\nonumber X: 220 \leq \parallel X-P_i \parallel_2 < 450, X \in D,& 0.83<\rho_{i,Query} \leq 0.9 \\
\nonumber X: 450 \leq \parallel X-P_i \parallel_2 < 900, X \in D, &  0.55<\rho_{i,Query} \leq 0.83 \\
X: 900 \leq \parallel X-P_i \parallel_2, X \in D,  & \rho_{i,Query} \leq 0.55.
\end{eqnarray}
\end{small}
\noindent The feasible region of localization can then be obtained by the intersection of the feasible regions obtained using all the anchor nodes as follows:
\begin{equation}
\label{eq:Circleintersection}
\widehat{P}_{Query}=\cap_{i}\widehat{P}_{i,Query} \hspace{4mm}, i \in\{1,2,\ldots ,K\}.
\end{equation}
\noindent We measure the performance of the proposed method in terms of probability of localization $p_{loc}$ and area of localization $a_{loc}$, as discussed in Section~\ref{subsec:ch6:halfplane} and defined in Eq.~(\ref{eq:PLoc}) and Eq.~(\ref{eq:ALoc}), respectively.

As we have ENF data from five locations, we use four locations as anchor cities and use the ENF data of the fifth city as the query to estimate its location using the correlation quantization relations described in Eq.~(\ref{eq:RhoDRelation}) and~(\ref{eq:AreaCorrelation}). In Figures~\ref{fig:Example5LocCircle}(a)--\ref{fig:Example5LocCircle}(d), we show the set of feasible area obtained after each constraint is added with College Park, Raleigh, Atlanta, and Champaign as anchor cities, and Princeton as the query city. In these figures, the shaded region represents the estimated feasible region for the query city. From this figure, we observe that as the number of constraints increases, the area of the feasible region decreases, thus improving the localization precisions. From Figure~\ref{fig:Example5LocCircle}(d) and Figure~\ref{fig:Example5Loc}(f), we observe that the localization precision using the correlation quantization method is better than the half-plane intersection method, as the value of $a_{loc}$ using the former method is less than the value of $a_{loc}$ for the latter method. The detailed comparison results between the two methods are discussed in the next section.

In our study, the correlation quantization levels are determined empirically. In general, the knowledge of grid layout and addition of anchor cities in the grid will introduce more constraints in the quantization scheme, and may provide a better and refined quantization scheme. An improved quantization scheme may lead to an improved localization accuracy measured in term of a reduction in feasible area of localization, $a_{loc}$, for a given probability of localization, $p_{loc}$. However, such a study requires availability of ENF databases from multiple locations throughout the grid. At this point of time, there are no such databases publicly available. Nevertheless, such a study can be conducted in future research to further rectify correlation coefficient-distance relationship to improve the localization accuracy.

\begin{figure*}
\begin{center}
\subfigure[$p_{loc}$-Princeton]{\includegraphics[width=.32\textwidth]{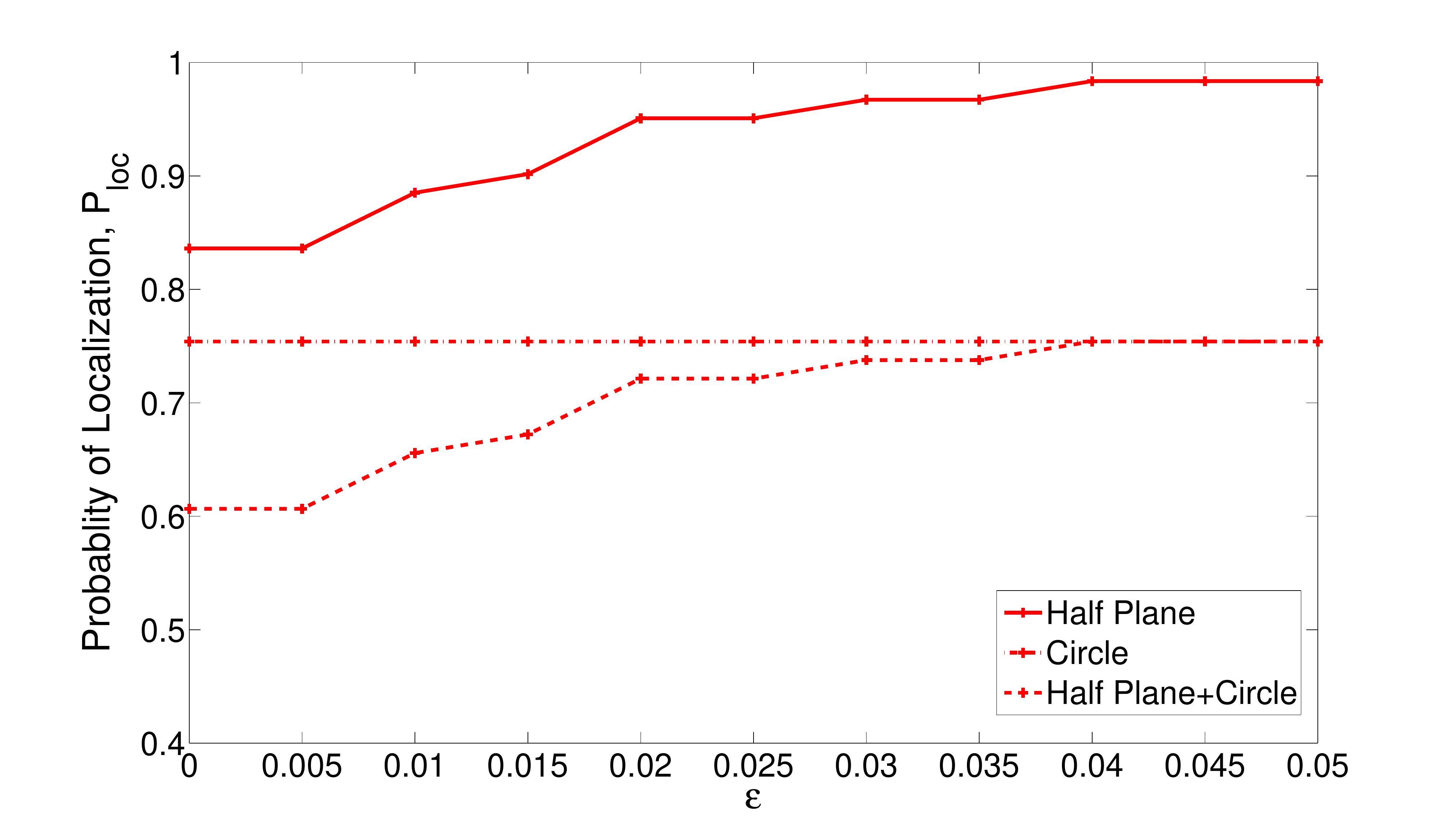}}
\subfigure[$p_{loc}$-College Park]{\includegraphics[width=.32\textwidth]{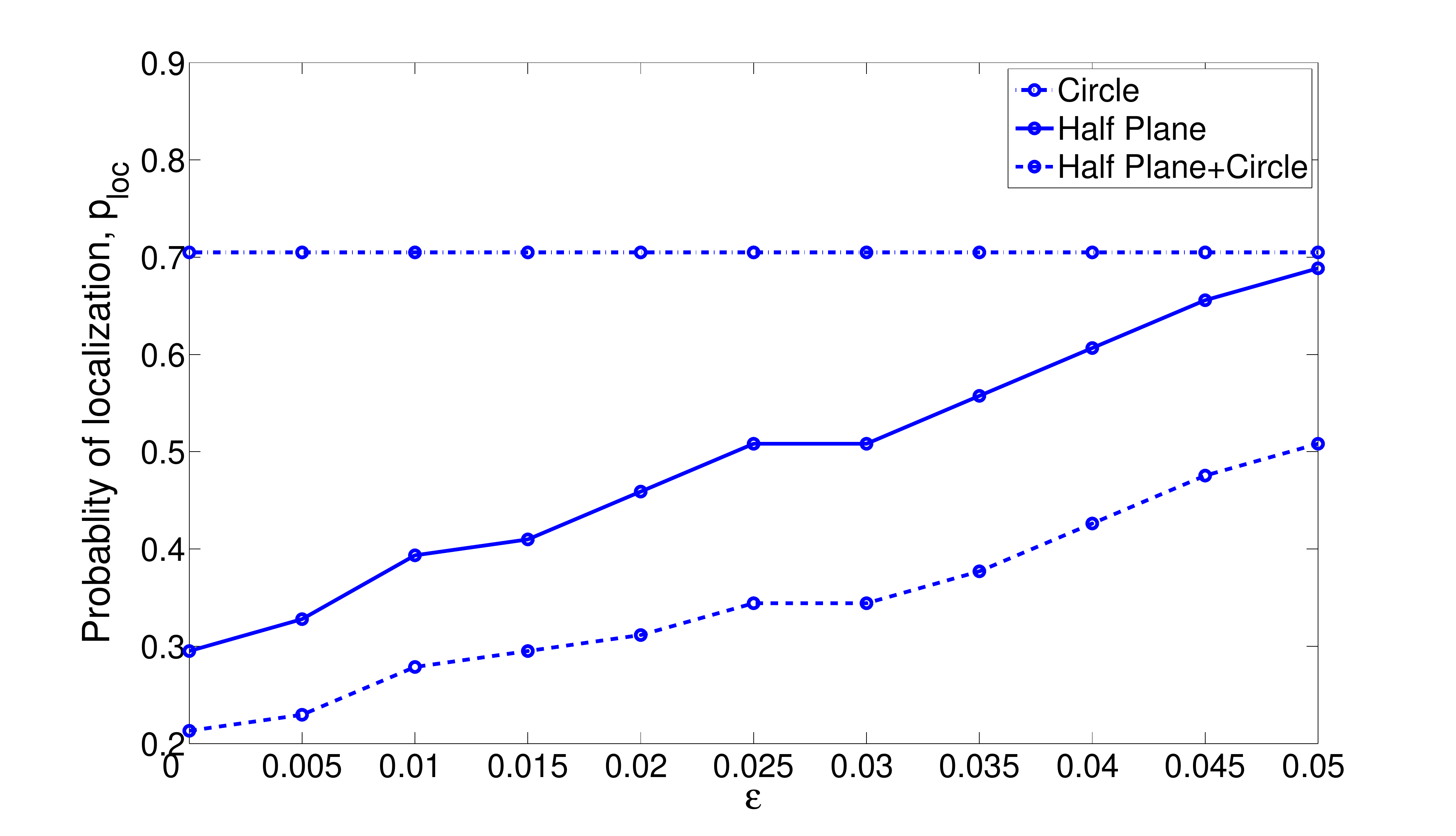}}
\subfigure[$p_{loc}$-Raleigh]{\includegraphics[width=.32\textwidth]{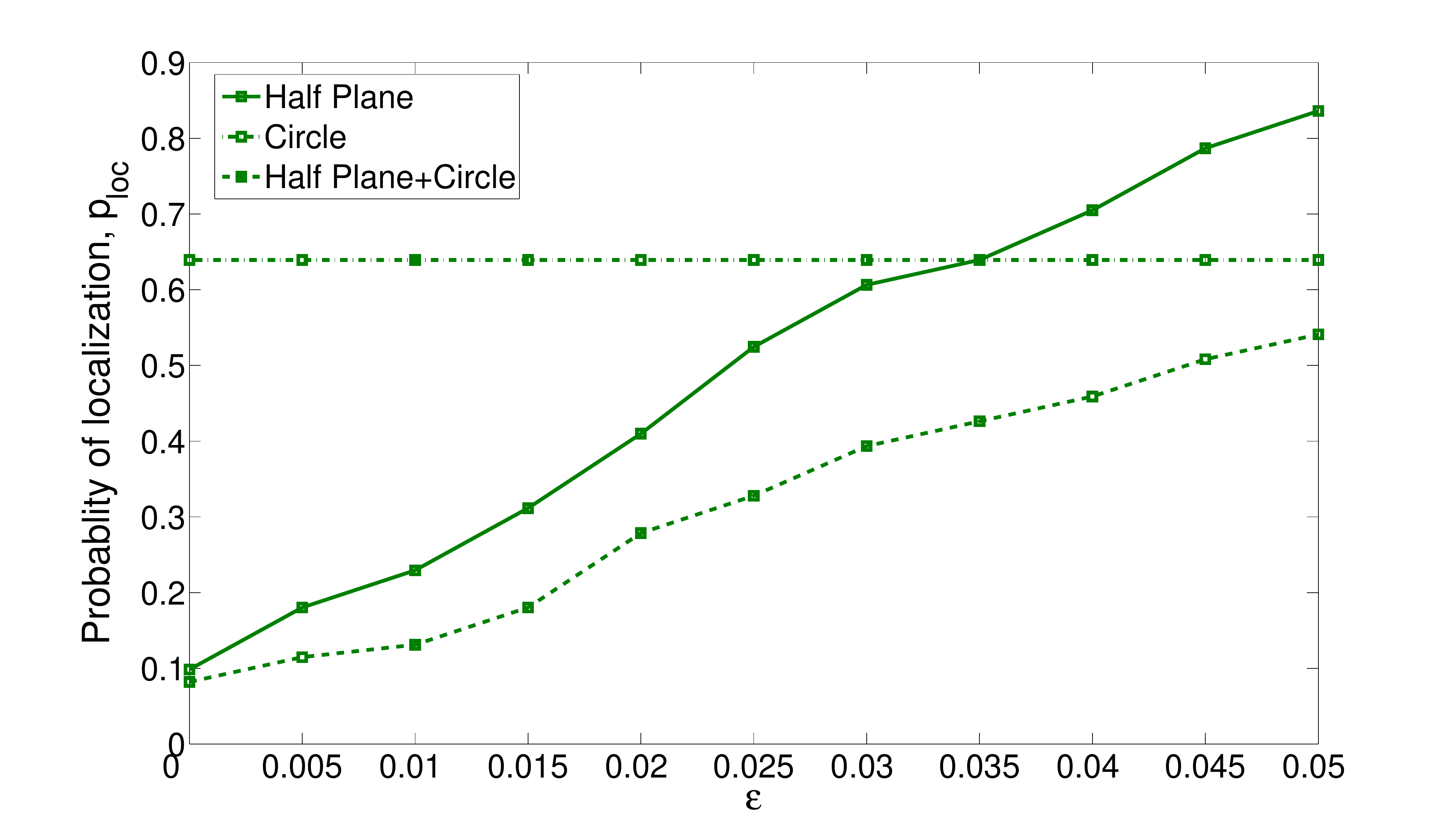}}
\subfigure[$a_{loc}$-Princeton]{\includegraphics[width=.32\textwidth]{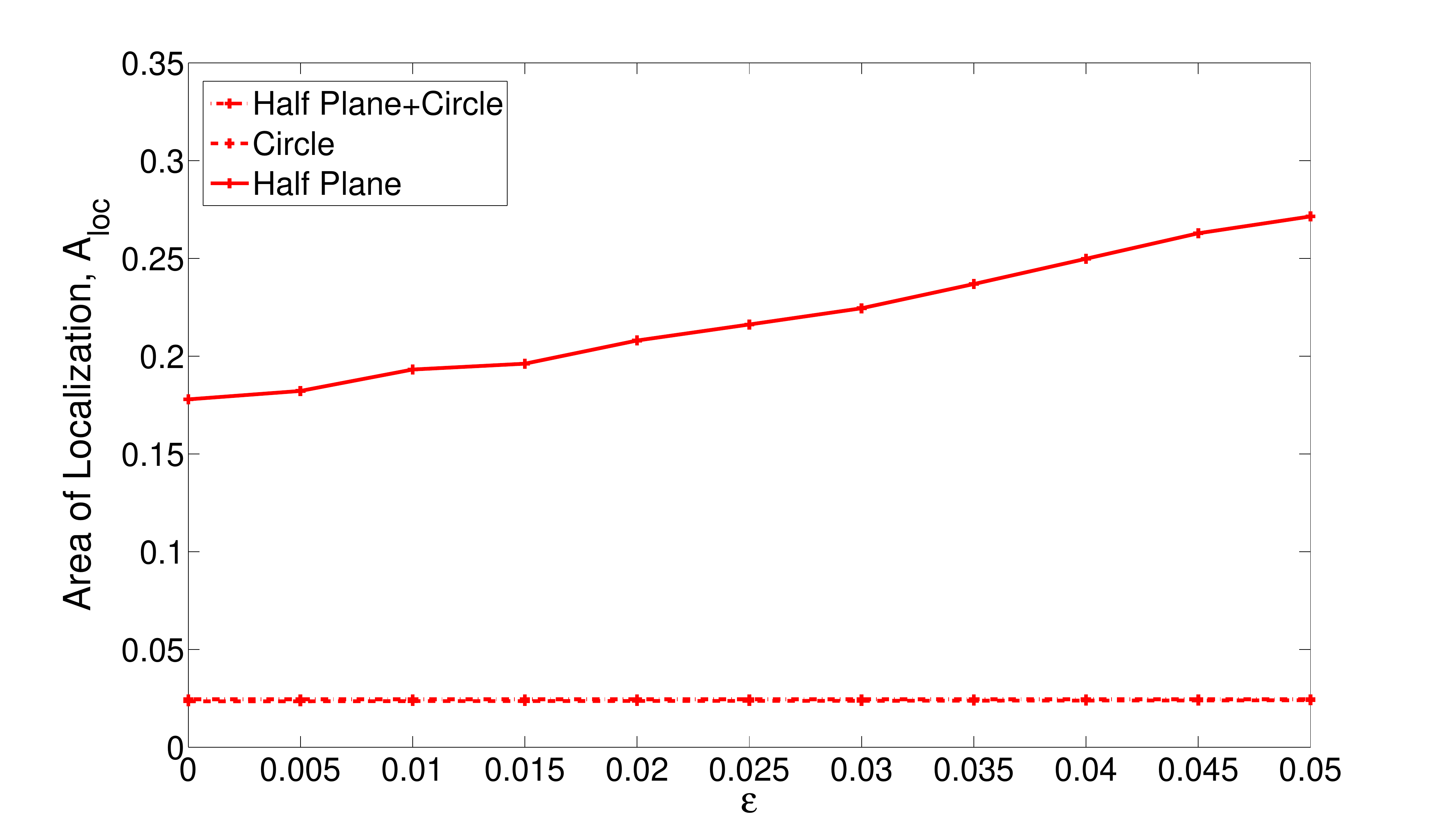}}
\subfigure[$a_{loc}$-College Park]{\includegraphics[width=.32\textwidth]{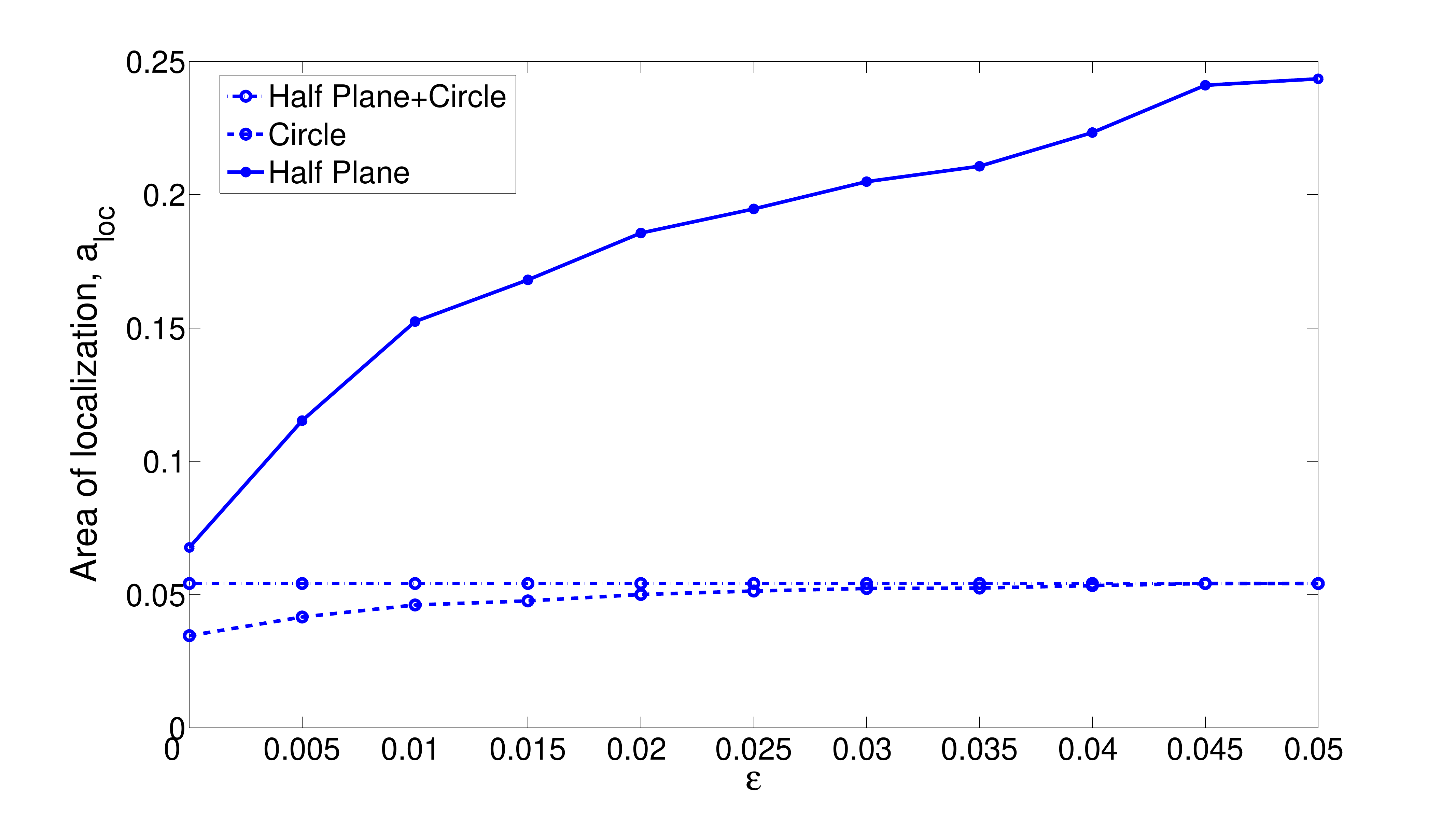}}
\subfigure[$a_{loc}$-Raleigh]{\includegraphics[width=.32\textwidth]{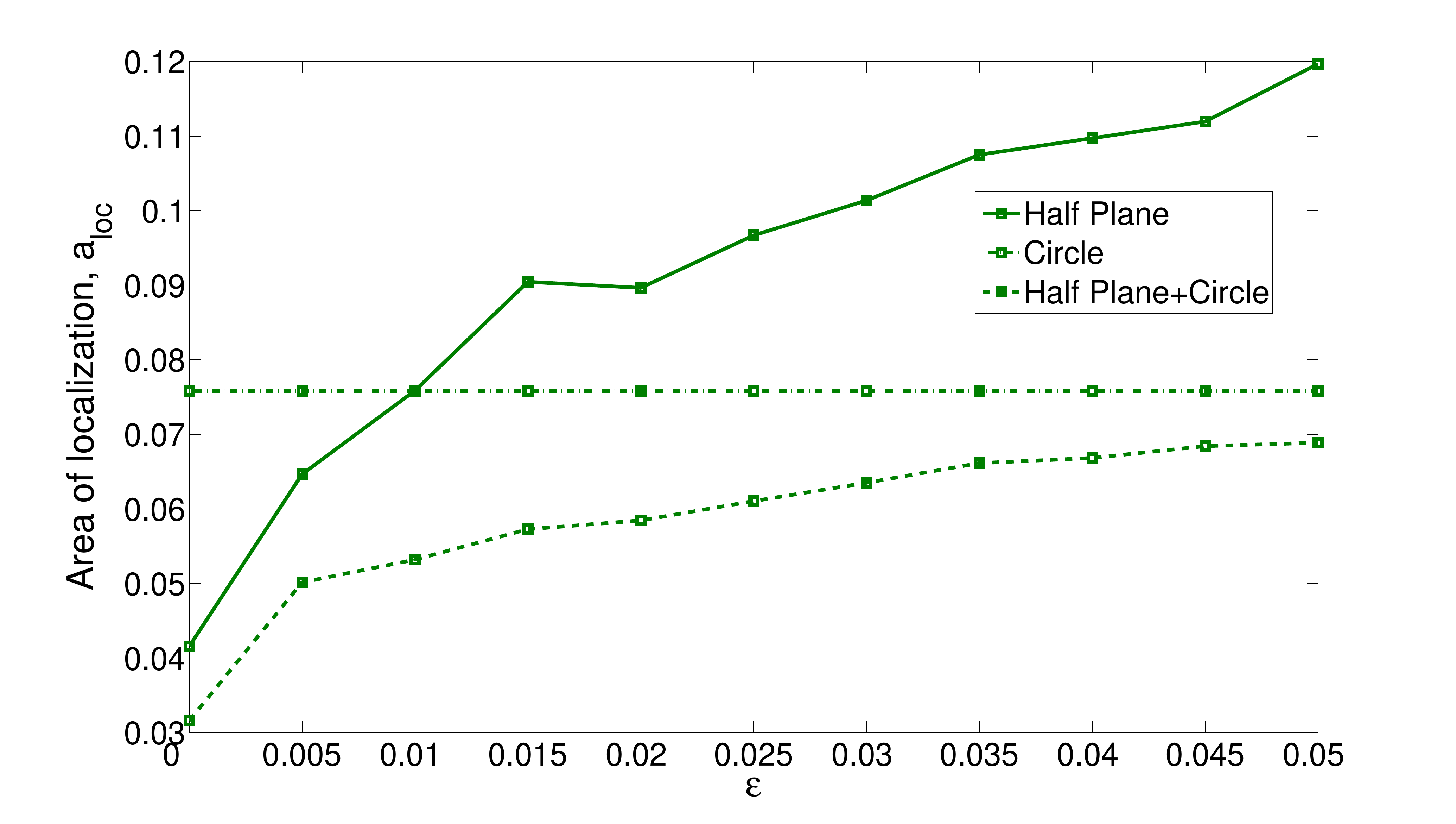}}
\subfigure[$p_{loc}$-Atlanta]{\includegraphics[width=.32\textwidth]{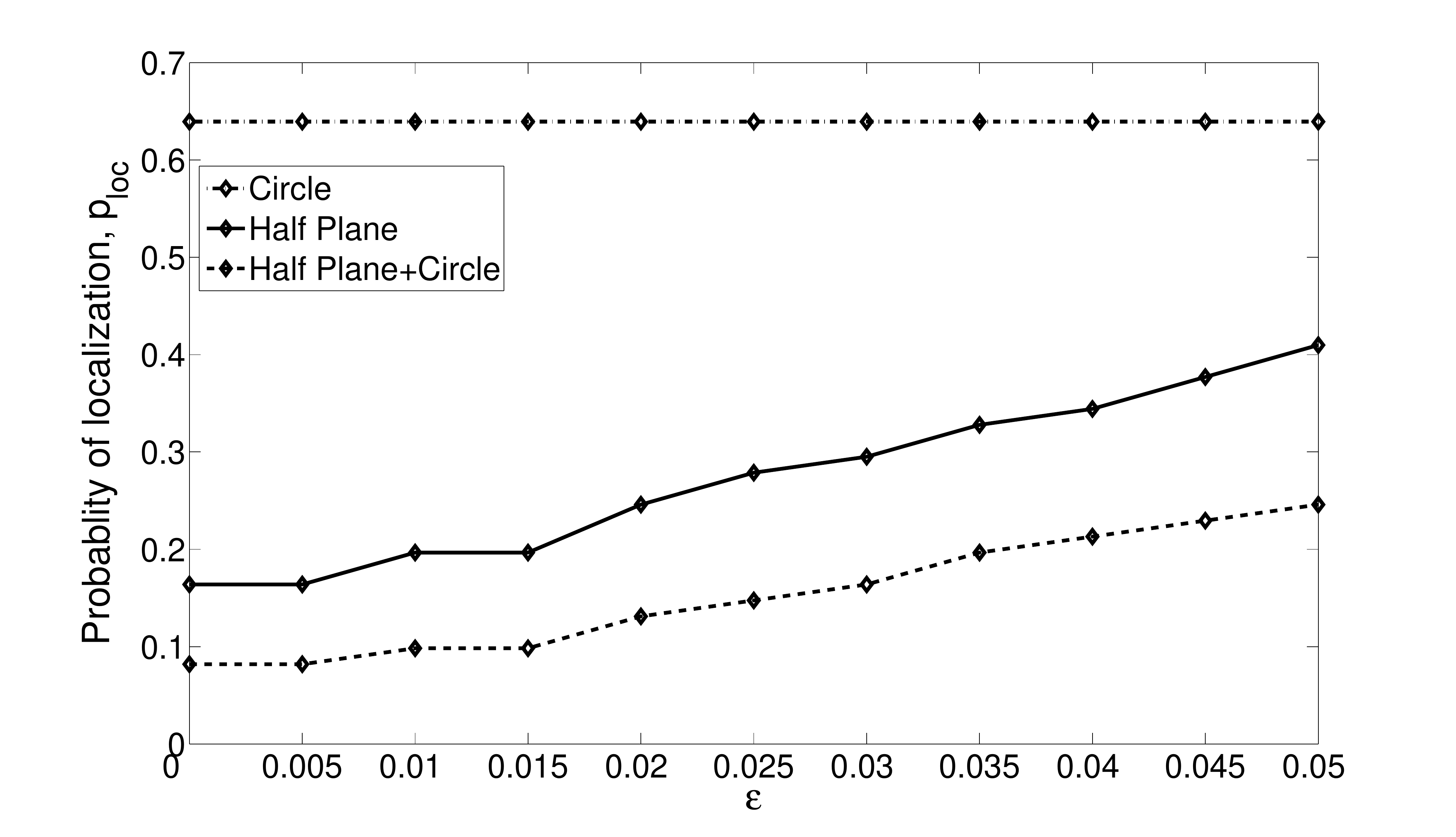}}
\subfigure[$a_{loc}$-Atlanta]{\includegraphics[width=.32\textwidth]{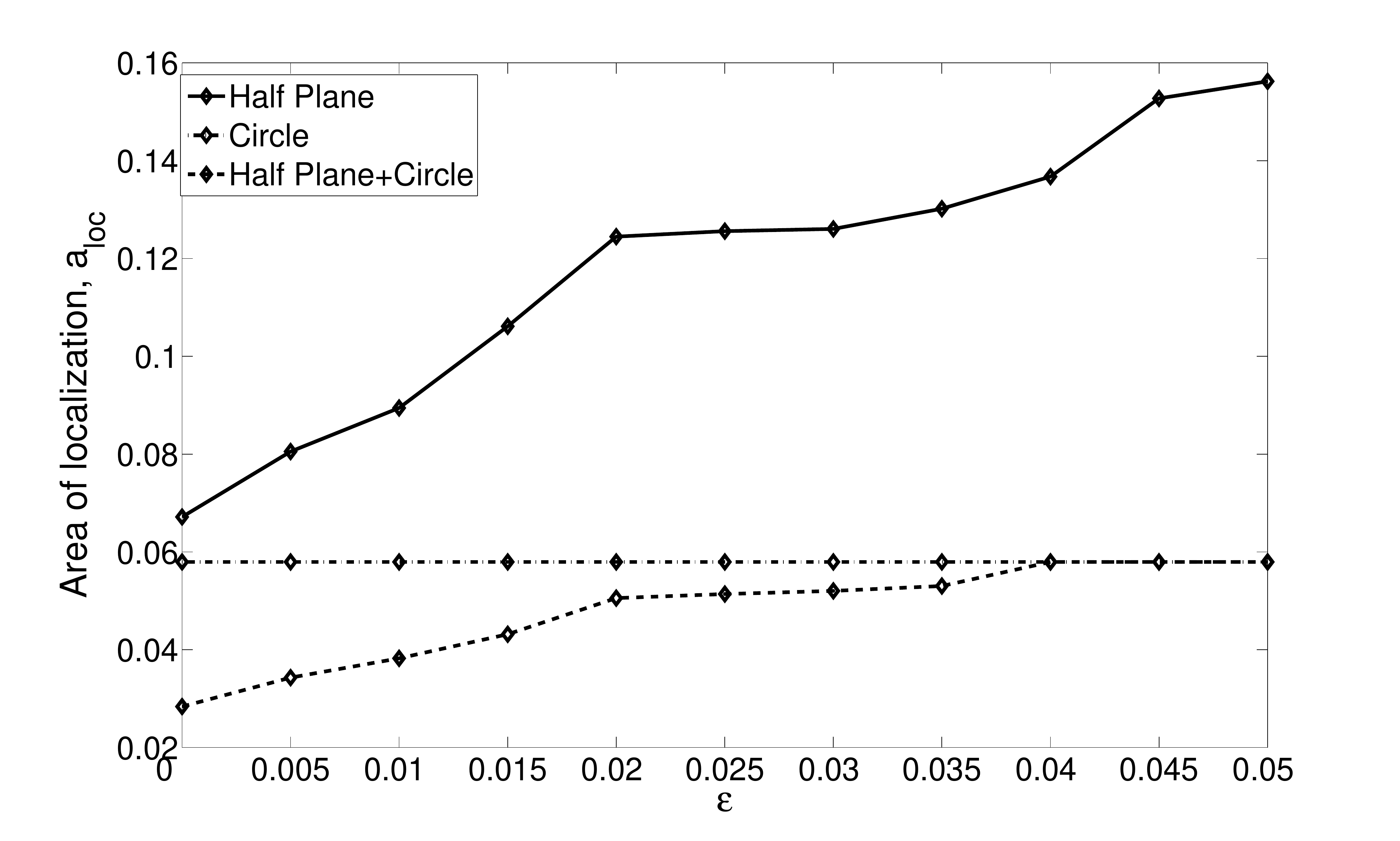}}
\end{center}
\caption[$p_{loc}$ and $a_{loc}$ using different localization methods]{$p_{loc}$ and $a_{loc}$ using different localization methods.}\label{fig:PLocALocUsingAllMethods}
\end{figure*}

\subsection{Combining the Two Methods}
We combine the localization constraints from the half-plane intersection method in Eq.~(\ref{eq:halfplaneNew}) and the correlation quantization method in Eq.~(\ref{eq:AreaCorrelation}) to obtain the localization accuracy. In Figure~\ref{fig:ExampleAll}(c), we plot an example of the localization feasible area using the combined method. The separated feasible areas of localization for the half-plane intersection method and the correlation quantization methods are shown in Figures~\ref{fig:ExampleAll}(a)--(b), respectively. From these figures, we observe that the combination of the two methods improves the estimation of the area of localization.

We plot the localization performance in terms of $p_{loc}$ and $a_{loc}$ of all three methods in Figures~\ref{fig:PLocALocUsingAllMethods}(a)--\ref{fig:PLocALocUsingAllMethods}(h) of the 5-location data with each query 8-minute long. From these figures, we observe that the localization precision of the correlation quantization method is better than that for the half-plane intersection method for all the cities for higher values of $\epsilon$, while the value of $p_{loc}$ for the half-plane intersection method may be slightly better as compared with that of the correlation quantization method. Using the constraints from the half-plane intersection method with the correlation quantization method slightly reduced the value of $p_{loc}$ for a given $\epsilon$ as compared with using the half-plane intersection method, but improves the localization precision by a significant amount. From these figures, it can be concluded that the combination of the half-plane intersection and the correlation quantization improves the localization performance over any method individually.

\subsection{Computational Complexity}
In this section, we discuss the computational complexity of the proposed localization methods. Both the methods proposed in this paper rely on estimating the pair-wise correlation coefficients between the temporally aligned ENF signals obtained from the query city and anchor cities. In the half-plane intersection method, correlation coefficients between ENF signals from the query city and two anchor cities are compared to obtain a half-plane feasible area. The maximum number of constraints that can be obtained for this method for $N$ anchor cities are $\binom{N}{2}$. Based on the sign of the difference between correlation coefficient from two anchor cities, the feasible area is divided into half-planes, as each constraint is added to the feasible solution.  So the computational complexity associated with the half-plane intersection method is $\mathcal{O}(N^2)$.

For the correlation quantization method, each pair-wise correlation coefficient provides a circular disc feasible region, with size of the disk dependent upon the quantization levels used for quantizing the correlation coefficient-distance relationship. Each anchor location contributes a constraint to solve for the feasible area of localization, thus making the computational complexity associated with $N$ anchor cities as $\mathcal{O}(N)$. However, this method requires a-priori knowledge of a sufficient number of data-points to quantize correlation coefficient-distance relationship, which are not needed in the half-plane intersection method.

\section{Further Discussions}
\label{sec:ch6:sensitivity}
In Section~\ref{subsec:ch6:3Locations} and~\ref{subsec:ch6:5LocationData}, we demonstrated the presence of location pin-pointing signatures in ENF signals. These results are obtained on ENF signals extracted directly from the power mains for the 3-location and the 5-location datasets. ENF signals extracted from power signals are clean signals with very high signal-to-noise ratio. Such quality is typically not observed in ENF signals embedded in multimedia recordings. It may be possible to use the analysis of Section~\ref{subsec:ch6:3Locations} and~\ref{subsec:ch6:5LocationData} in multimedia forensics, when the recording device pro-actively captures the ENF signal from the power mains and embeds it into the multimedia recording. Such scenarios are feasible by adding a power capturing device to audio recorders and cameras, and conducting the recording using the devices while connected to the power mains. In practical scenarios, only anchor-node recordings can possibly be acquired this way.

\begin{figure*}
\begin{center}
\subfigure[Query city=Princeton]{\includegraphics[width=.32\textwidth]{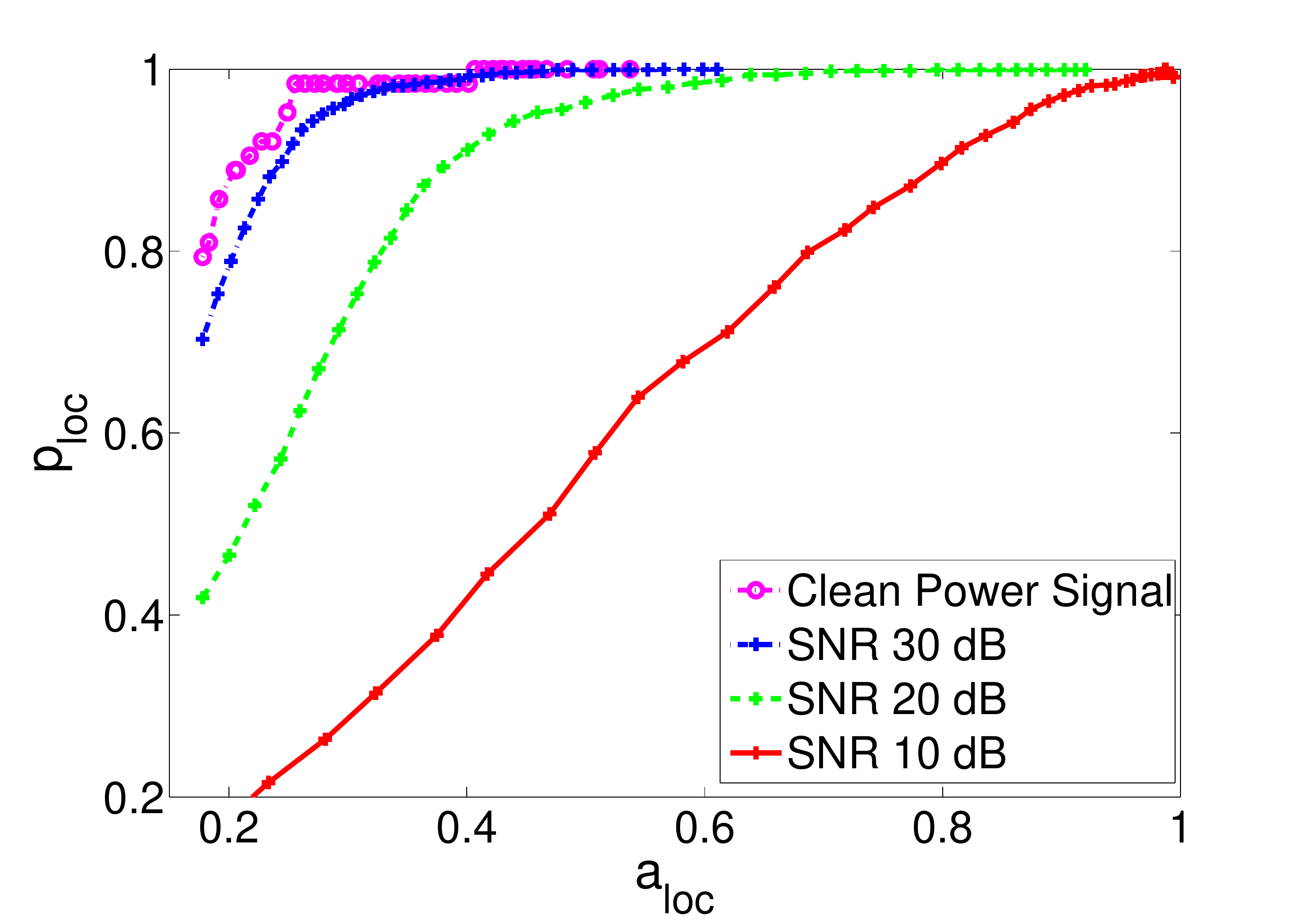}}
\subfigure[Query city=College Park]{\includegraphics[width=.32\textwidth]{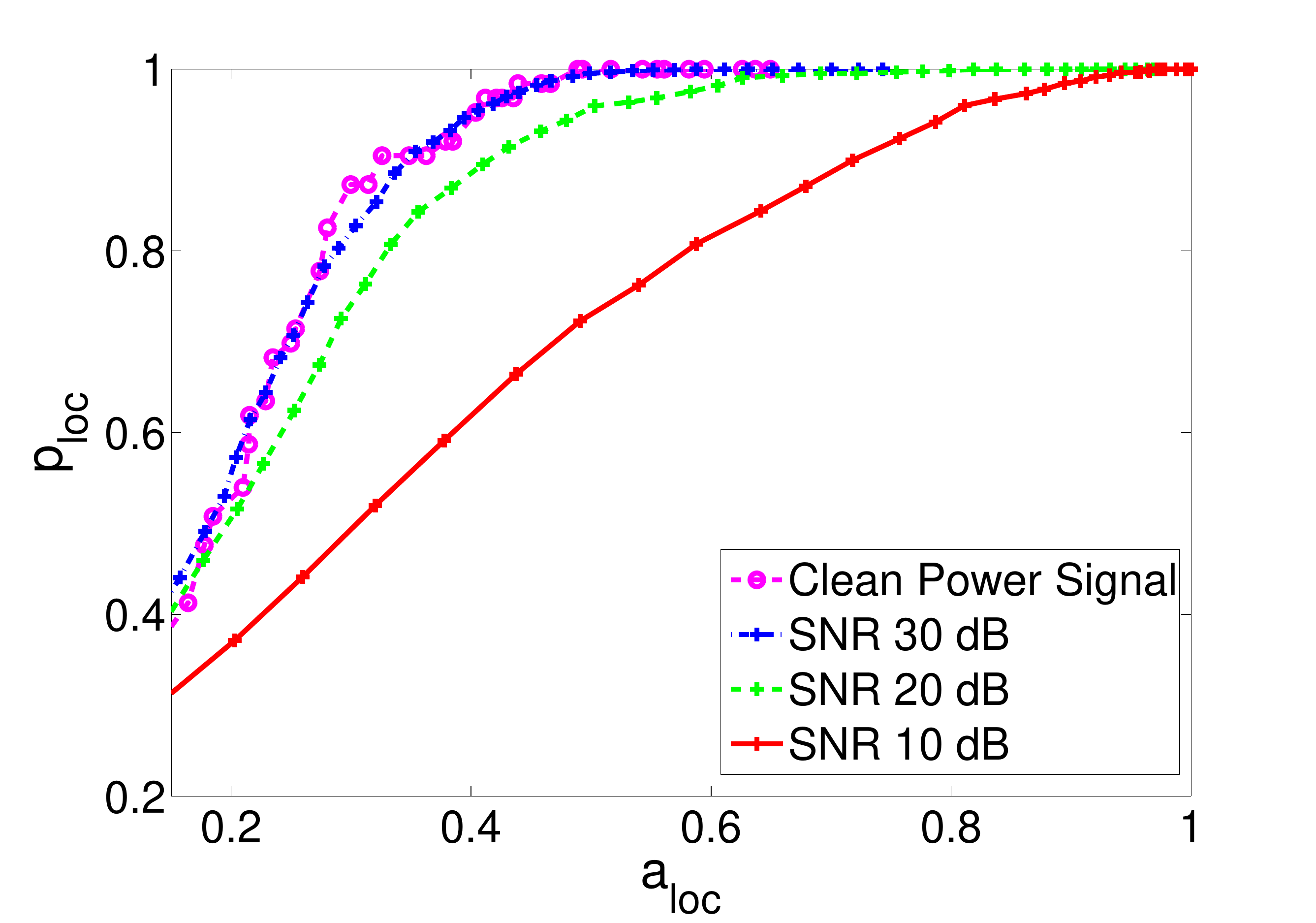}}
\subfigure[Query city=Raleigh]{\includegraphics[width=.32\textwidth]{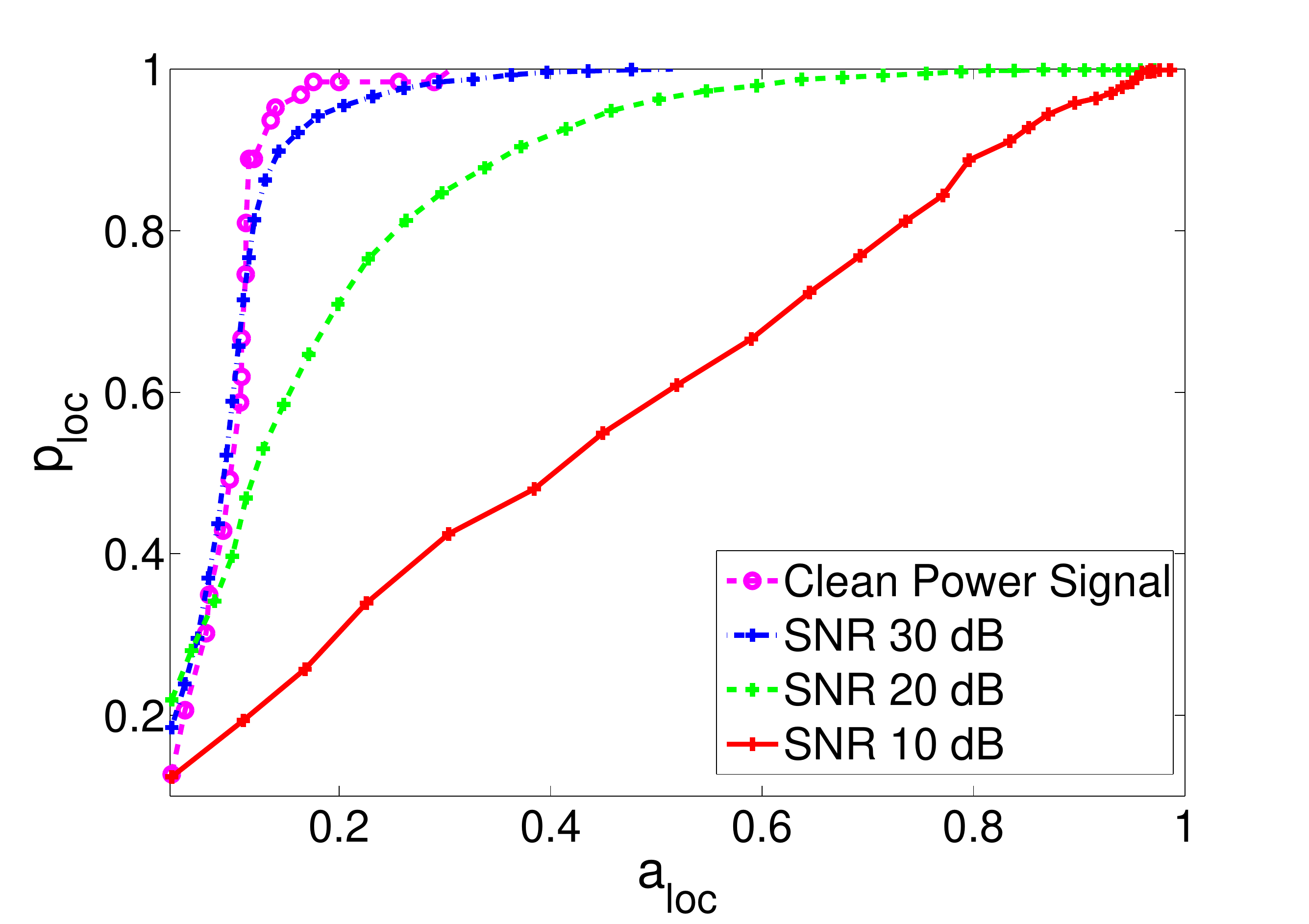}}
\end{center}
\caption[$p_{loc}$ under different noisy conditions using half-plane intersection method]{Probability of localization $p_{loc}$ under different noisy conditions for 5-location data using half-plane intersection method.}
\label{fig:NoiseSensitivityPLoc}
\end{figure*}

\paragraph{Effect of noise on localization} To understand the capabilities and limitations of ENF signal analysis for intra-grid location estimation from multimedia recordings, we conduct a study on the effect of noise on the localization capabilities of ENF signals. We evaluate the localization performance of the half-plane intersection method under the assumption that the ENF signal extracted from the query data is submerged in additive white Gaussian noise. This scenario is representative of cases where the ENF signals at anchor nodes are obtained from the power mains recordings, and the query data are multimedia recordings capturing ENF traces. We evaluate the effect of different noise levels in the query ENF signal, and plot $a_{loc}$ v.s. $p_{loc}$ in Figures~\ref{fig:NoiseSensitivityPLoc}(a)--\ref{fig:NoiseSensitivityPLoc}(c) for Princeton, College Park, and Raleigh as query cities, respectively. From these figures, we observe that the localization performance for all three query cities improves with an increase in signal-to-noise ratio (SNR). For SNR$>$30~dB, the localization performance approaches the localization performance obtained using the clean power ENF signal as the query signal. SNR$>$20~dB is required to achieve a reasonable localization accuracy in media recordings with some practical applications.

\paragraph{Challenges with multimedia signals} The ENF estimates obtained from a power signal can be considered its cleanest, most readily available form. Following this, given two simultaneously recorded signals, a power signal and an audio signal, an estimate of the SNR for the audio ENF signal can be obtained by subtracting the audio ENF signal from the power ENF signal. The value of SNR for ENF signals extracted from an audio recording used in~\cite{ENFModel} was estimated 15~dB when instantaneous ENF was estimated using a frame of 16-second, and 9~dB when estimated using 8-second long frame~\cite{Ravi_Thesis_Drum}. The SNR of audio ENF signals reduces further on decreasing the frame size to 1-second, the optimal frame size to achieve a good localization accuracy, as shown in our experiments in Sec.~\ref{subsec:ch6:3Locations}. The noise sensitivity analysis clearly indicates that an SNR of more than 20~dB in the query signal is necessary for a good localization performance. Based on these observations, it is difficult to utilize the ENF signals for intra-grid localization in multimedia recordings using the state-of-the-art ENF estimation techniques, and localization feature extraction techniques presented in this paper. A breakthrough in ENF estimation techniques is needed to extract a high SNR ENF signal from media recordings. Nevertheless, this study demonstrates that intra-grid location specific signatures are present in ENF signals, which can be used to design scalable localization methods.

\paragraph{Emerging applications of ENF based localization} As our research progresses, we see that fine localization signatures in ENF has applications beyond audio or video forensics, most notably in establishing secure connected autonomous IoT and cyber physical systems (CPS)~\cite{Wu:2017:Sensys}. A potential use-case model is using ENF traces from speciality designed sensors deployed in IoT/CPS applications to help authenticate location, by providing location authentication in security tokens, or providing ``Proof-Carrying Sensing'' for cyber physical systems. For example, deployment of IoT/CPS applications can have each connected device equipped with such sensors as photo-diodes with a frequency response encompassing ENF bands of interest for indoor sensing~\cite{Garg}, speciality designed acoustics sensors to capture ENF at high SNR, or sensors to record ENF from power-mains connected devices. High SNR ENF from these devices can then be used for location authentication by these devices after verifying the claimed location with that of the estimated location using ENF signal from the device and the anchor nodes in the grid.

\section{Conclusions and Future Work}
\label{sec:ch6:discuss}
In this paper, we have discovered ``location signatures'' present in the high frequency details of ENF signals. We have demonstrated a linear relationship between the correlations of temporally aligned location signature signals at different locations and geographical distances between the locations in the densely populated US east and west power grids. We have harnessed such a relationship to propose two trilateration based localization approaches, the half-plane intersection method and the correlation quantization method, to estimate the location at which recordings were made, without any need of a reference or training recording obtained from the query location. A combination of the half-plane intersection method and the correlation coefficient quantization method provided the best localization performance measured in terms of the probability of localization, $p_{loc}$, and the area of localization, $a_{loc}$, defined in this paper. The distance-correlation relationship can be further rectified and the localization accuracy improved by adding more locations as anchor nodes, thus providing a scalable method for finer localization.

The focus of this paper has been on exploring the unchartered application of ENF signal analysis for intra-grid location estimation of multimedia data. This first study conducts experiments on power ENF signals and provides encouraging results in that direction. Multimedia ENF data are, however, more challenging than power ENF data due to the presence of noise. As we employ the high frequency variations of the ENF signal to extract a meaningful metric for localization, the noisy nature of the ENF signal in multimedia data may increase the difficulty in localization. Furthermore, as shown by our experiments, location specific variations are best captured using instantaneous frequencies estimated at a one-second temporal resolution; reliable ENF signal extraction from multimedia data at such a high temporal resolution also presents a research challenge.

The results presented in this paper demonstrate that ENF signals offer a strong potential to be used as a location-stamp. In future work, we plan on exploring approaches that would allow us to extend the localization work based on power ENF signals to audio and video ENF signals. For that purpose,  we need to gain a stronger understanding on the conditions and factors that play a role in promoting or hindering capturing of ENF traces in media recordings~\cite{Adi_Thesis_Drum}. We also need to explore methods to improve accuracy and robustness of ENF signal extraction with a high temporal resolution and under noisy conditions, more akin to typical audio and video recordings.

{\small
\bibliographystyle{ieeetran}
\bibliography{Fine_location_JOSP_Arxiv}
}

\end{document}